\documentclass[useAMS]{mn2e}
\usepackage{graphicx}
\usepackage{amssymb}
\usepackage{footnote}
\usepackage{threeparttable}
\usepackage{epsfig}
\usepackage{pslatex}
\usepackage{amsmath}
\usepackage{epstopdf}

\title[Recurrent nova RS Oph]{Optical spectroscopy of the recurrent nova RS Ophiuchi - from the outburst of 2006 to quiescence}

\author[Mondal et al.]{Anindita Mondal,$^{1}$
G. C. Anupama,$^{2}$\thanks{E-mail: gca@iiap.res.in (GCA)}
U. S. Kamath,$^{2}$
Ramkrishna Das,$^{1}$
\newauthor G. Selvakumar$^{3}$
and Soumen Mondal$^{1}$\\ \\
$^{1}$S N Bose National Centre for Basic Sciences, Salt Lake, Kolkata 700 106, India\\
$^{2}$Indian Institute of Astrophysics, II Block Kormangala, Bangalore 560 034, India\\
$^{3}$Vainu Bappu Observatory, Indian Institute of Astrophysics, Kavalur, Alangayam 635 701, India\\
}

\begin{document}

\pagerange{\pageref{firstpage}--\pageref{lastpage}} \pubyear{2017}
\maketitle

\begin{abstract}
Optical spectra of the 2006 outburst of RS Ophiuchi beginning one day after discovery to over a year after the outburst are presented here.
The spectral evolution is found to be similar to that in previous outbursts. The early phase spectra are dominated by hydrogen and helium (I \& II) lines. Coronal and nebular lines appear in the later phases. Emission line widths are found to narrow with time, which is interpreted as a shock expanding into the red giant wind.  Using the photoionisation code CLOUDY, spectra at nine epochs spanning 14 months after the outburst peak, thus covering a broad range of ionisation and excitation levels in the ejecta, are modelled. The best-fit model parameters indicate the presence of a hot white dwarf source with a roughly constant luminosity of 1.26 $\times$ 10$^{37}$ erg~s$^{-1}$. During first three months, the abundances (by number) of He, N, O, Ne, Ar, Fe, Ca, S and Ni are found above solar abundances; abundances of these elements decreased in the later phase. Also presented are spectra obtained during quiescence. Photoionisation model of the quiescence spectrum indicates the presence of a low luminosity accretion disk. The helium abundance is found to be subsolar at quiescence.
\end{abstract}


\section{Introduction}
\label{sect:intro}

RS Ophiuchi (RS Oph) is a recurrent nova that has had nova outbursts in 1898, 1933, 1958, 1967 and 1985 (Rosino 1987), with possible
outbursts in 1907 (Schaeffer 2004) and 1945 (Oppenheimer \& Mattei 1993). It is a fast nova with $t_3=8-10$~days, and the
optical spectrum is characterized by strong coronal and other high excitation lines that reach a peak around 60 days after maximum (e.g. Rosino 1987; Anupama \& Prabhu 1989). The outbursts occur due to thermonuclear runaway (TNR) on a massive white dwarf (WD) surface that has accreted matter from its red giant companion star (Starrfield et al. 1985; Yaron et al. 2005). The fast-moving nova ejecta interact with the slow-moving red giant wind, leading to the formation of shocks (Bode \& Kahn 1985, O'Brien and Kahn 1987). This leads to the formation of coronal lines, and also the narrowing of the emission lines (e.g. Gorbatskii 1972, 1973; Snijders 1987; Anupama \& Prabhu 1989; Shore et al. 1996). There is a remarkable similarity between the optical light curves and spectra from different outbursts (Rosino 1987). So far, the 1985 outburst was one of the best studied events, being recorded from X-rays to radio wavelengths (Bode 1987 and references therein).

The sixth recorded outburst of the recurrent nova RS Oph was discovered on 2006 February 12.83 (Narumi et. al., 2006). It offered another chance for
multiwavelength observations, and has been studied in detail at several wavelength regimes; e.g. in X-rays (e.g. Bode
et al. 2006; Sokoloski et al. 2006), in the optical (Buil 2006; Fujii 2006; Iijima 2006; Skopal et al. 2008), in the infrared (IR) (e.g. Das et al.
2006; Banerjee et al. 2009; Monnier et al. 2006; Evans et al. 2007) and in the radio (e.g. O'Brien et al. 2006; Kantharia et al. 2007;
Rupen et al. 2008). The X-ray results clearly detect an X-ray blast wave that expands into the red giant wind. The temporal
evolution of the shock wave was traced from XRT observations from the Swift satellite (Bode et al. 2006)
and the Rossi X-ray Timing Explorer (RXTE) observations (Sokoloski et al. 2006). A similar shock has been detected from the evolution
of the line widths in the near-IR spectra of the 2006 outburst (Das et al. 2006). The 2006 outburst also showed similarity to the previous outburst,
with some differences, especially in the X-rays and radio.

In this paper, we present the evolution of the optical spectrum of RS Oph during 14 months following the 2006 outburst based on spectra obtained using
the 2m Himalayan Chandra Telescope (HCT), Hanle, India and the 2.3m Vainu Bappu Telescope (VBT), Kavalur, India. We have also modelled the observed spectra with the photoionisation code CLOUDY. From the best fit spectra, we have estimated physical parameters such as luminosity, temperature, elemental abundances etc. associated with the RS Oph system.

\section{Observations and data reduction}
\label{sect:obsns}

The observations of RS Oph began on 2006 February 13, one day after the
outburst discovery and continued well after the nova had reached quiescence.
Low resolution spectra were obtained using the HCT and occasionally with the
VBT, while high resolution (R$\sim 30,000$) spectra were obtained with the VBT
in the immediate post-maximum phase (14 and 19 February 2006). The HCT spectra 
were obtained using the Himalaya Faint Object Spectrograph Camera (HFOSC) 
through Grism \#7 {(R$\sim$ 1500)} and Grism \#8 {(R$\sim$ 2200)}, covering the wavelength ranges of 
3400-8000 \AA\ and
5200-9200 \AA\ respectively. The high resolution VBT spectra were obtained covering the wavelength ranges of 4000 - 8000 \AA\, using
the fiber-fed Echelle spectrograph. Table 1
gives the dates of observations. We adopt the date of maximum as 2006 February 12.83 (JD 2453779.330).\\

All spectra were reduced in the standard manner using the various tasks under the IRAF package. The spectra were bias subtracted, flat field corrected and the spectrum extracted using the optimal extraction method.  The HFOSC spectra were calibrated using the FeAr calibration source for the blue region, and FeNe calibration source for the red region. Spectroscopic standards obtained during the same night were used for instrumental response correction. The blue and the red spectra of the same night were combined and spectra obtained on a relative flux scale, which were then flux calibrated using the photometric magnitudes published in the literature. All spectra were then de-reddened using $E(B-V)=0.71$.
The VBT Echelle spectra were wavelength calibrated using the ThAr source. These spectra are not flux calibrated.

\begin{table}
\label{jnl_obs}
\centering
\begin{minipage}[]{100mm}
\caption[]{Journal of observations of RS Oph during 2006.}
\end{minipage}
\setlength{\tabcolsep}{1pt}
\small
\begin{tabular}{lcr}
\hline\noalign{\smallskip}
Date & Telescope & Instrument \\
\hline\noalign{\smallskip}
13 Feb & HCT & HFOSC \\
14 Feb & VBT & Echelle \\
15 Feb & HCT & HFOSC \\
16 Feb & HCT & HFOSC \\
17 Feb & HCT & HFOSC \\
18 Feb & HCT & HFOSC \\
19 Feb & VBT & Echelle \\
20 Feb & HCT & HFOSC \\
21 Feb & HCT & HFOSC \\
24 Feb & HCT & HFOSC \\
26 Feb & HCT & HFOSC \\
27 Feb & HCT & HFOSC \\
28 Feb & HCT & HFOSC \\
2 Mar & HCT & HFOSC \\
10 Mar & HCT & HFOSC \\
12 Mar & HCT & HFOSC \\
14 Mar & HCT & HFOSC \\
16 Mar & HCT & HFOSC \\
21 Mar & HCT & HFOSC \\
24 Mar & HCT & HFOSC \\
1 Apr & HCT & HFOSC \\
3 Apr & HCT & HFOSC \\
24 Apr & HCT & HFOSC \\
26 Apr & HCT & HFOSC \\
3 May & HCT & HFOSC \\
26 May & HCT & HFOSC \\
1 Jun & HCT & HFOSC \\
1 Jul & HCT & HFOSC \\
4 Jul & HCT & HFOSC \\
19 Jul & HCT & HFOSC \\
18 Aug & HCT & HFOSC \\
17 Oct & HCT & HFOSC \\
\hline\noalign{\smallskip}
\end{tabular}
\end{table}

\section{Evolution of the optical spectrum}
\label{sect:evoln}

\subsection{The early phase : days 1 to 15}
\label{subsect:earlyphase}
One day after the maximum, the spectrum is dominated by broad emission
lines due to H, N, Ca II (IR), Na I and O I. Weak P Cygni
absorption features are seen associated with the emission lines.  The nova appears to be in the ``fireball expansion" phase.
About 5 days after maximum, P Cyg absorption features associated with the broad emission are absent and
the emission line widths are narrower. The continuum appears bluer. The prominent lines are due to H and Fe II. [Ar X] 5535 line could be weakly present. The strength of O I 8446
line has increased while the strength of Ca II IR lines has decreased.  [Fe X] 6374 begins to appear around day 8. Sharp [O III] 4959, 5007, [N II] 5755
appear around day 15. [Fe XI] is probably present on day 15. These lines most probably arise from the
ionised red giant wind. In the spectra of days 14-16, the He I lines show a 
double peaked profile. The spectra are shown in Figures 1 and 2.

RS Oph was detected in the 14-25 keV band of {\it{Swift}} XRT and in the 3-5 keV
and 5-12 keV bands of the {\it{RXTE}} all sky monitor during $\sim 0-5$ days, 
with an increasing flux (Bode et al. 2006, Sokoloski et al. 2006). The presence 
of the coronal lines in the optical spectrum during the very early phase appears
 to be consistent with the hard X-ray detection. The increase in the blue 
continuum is also consistent with the increase in the X-ray flux.

The high resolution spectra of day 1 show sharp, narrow P Cygni emission 
features superimposed over the broad emission features (see 
Figure 3). The P Cyg absorption, which originates in the 
circumstellar red giant wind indicates a velocity $\sim 20$ km s$^{-1}$. The 
absorption decreased in strength around day 7, similar to the broad absorption 
component. The wind emission is dominated by Fe II emission lines. The spectra 
clearly show the Na ID interstellar absorption components as well as several 
diffuse interstellar bands between 5690 \AA\ and 5870 \AA\, at 5705, 5778, 5780,
 5797, 5844 and 5849 ~\AA\ (Josafatsson \& Snow 1987). Of these, the narrow 
bands at 5780, 5797 and 5849 \AA\ are correlated with interstellar reddening. 
Using the average of the 5797 \AA\ narrow band equivalent width of 
$0.109\pm 0.01$ \AA, an $E(B-V)=0.75\pm 0.05$ is estimated.

\begin{figure}
\resizebox{\hsize}{!}{\includegraphics{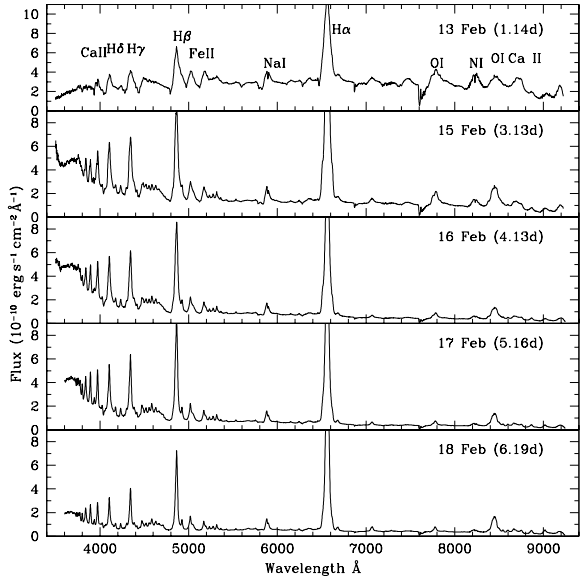}}
\caption[]{Spectra of RS Oph during 2006 February. The most prominent features are those due to  H, Fe II and Ca II. H$\alpha$ line has been truncated so that the weaker lines are seen properly.}
\label{specfeb1}
\end{figure}

\begin{figure}
\resizebox{\hsize}{!}{\includegraphics{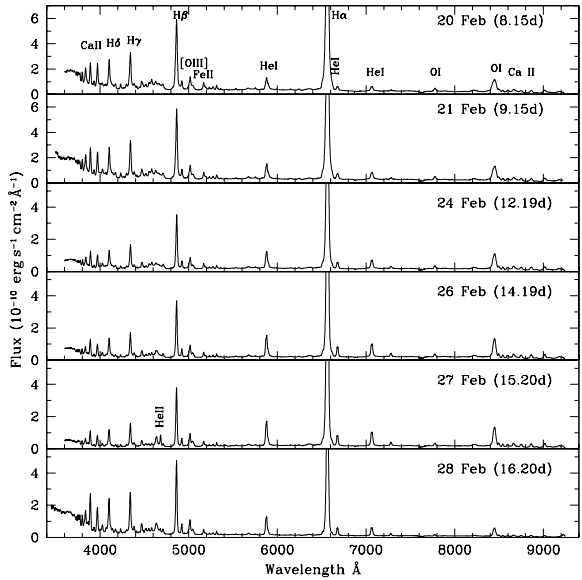}}
\caption[]{Spectra of RS Oph during 2006 February (continued). Note the emergence and strengthening of the He II 4684~\AA\ line during days 8-15, and the sudden decline on day 16.}
\label{specfeb2}
\end{figure}

\begin{figure}
\resizebox{\hsize}{!}{\includegraphics{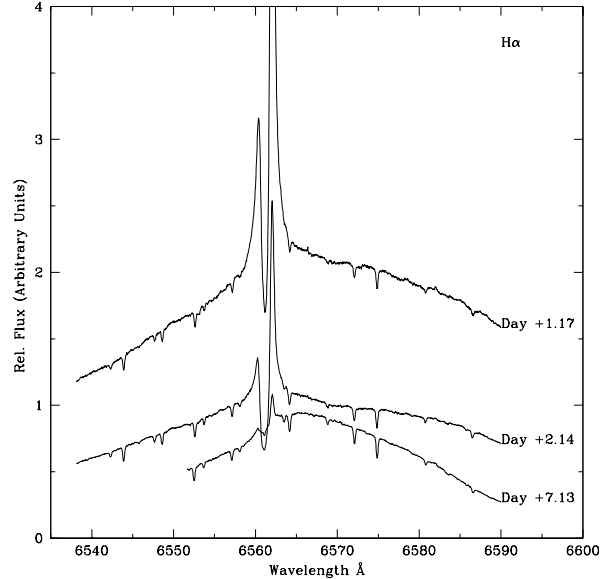}}
\caption[]{The H$\alpha$ line in the high resolution spectra during days 1-7. The emission from the nova ejecta is broader than the wavelength covered by the order. The plot shows the narrow P Cygni component from the circumstellar material. Note the decrease in the strength of this
component by day 7.}
\label{halpha}
\end{figure}

\begin{figure}
\resizebox{\hsize}{!}{\includegraphics{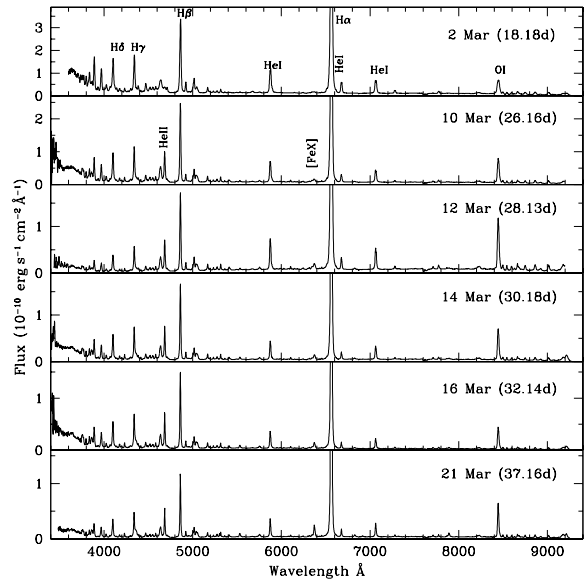}}
\caption[]{Spectra of RS Oph during 2006 March. Fe II lines have weakened, and the lines due to He have strengthened. Also note strengthening of coronal lines and OI 8446~\AA.}
\label{specmar}
\end{figure}

\subsection{The high excitation phase : days 16 to 80 }
\label{subsect:hiexcit}

This phase is marked by a decrease in the strength of the permitted lines, and the strengthening of high excitation lines. The Fe II lines fade, while the
He I lines develop and strengthen with time, and by $\sim$ day 25, the spectrum is dominated by H and He (I \& II) lines (Figure 4).
Also, O I 8446 becomes stronger than O I 7774. The emergence of the
He II lines is consistent with the onset of the supersoft X-ray phase (SSS phase) that commenced about 30 days since outburst (Osborne et al. 2011).
Although a few coronal lines began developing as early as day 8, the onset of
the coronal phase appears to be around day 25, with all the coronal lines, due to [Fe X], [Fe XI] and [Fe XIV] strongly present, and strengthening with
time. The coronal phase peaks around 60-70 days after maximum, similar to the previous outbursts (Anupama \& Prabhu 1989).

The line
profiles show multiple components. The width of the emission lines continues to decrease with time. At around after 70 days, [O III] line width becomes
similar to those of other lines, and the  spectrum is dominated by Balmer, He II, coronal and other high excitation lines. O I 8446  shows a two
component -- a broad and a narrow component -- structure. By day 80, the [O III] lines are strengthening, and the O I 8446 \AA\
two component structure continues. The spectra are shown in Figure 5.

\begin{figure}
\resizebox{\hsize}{!}{\includegraphics{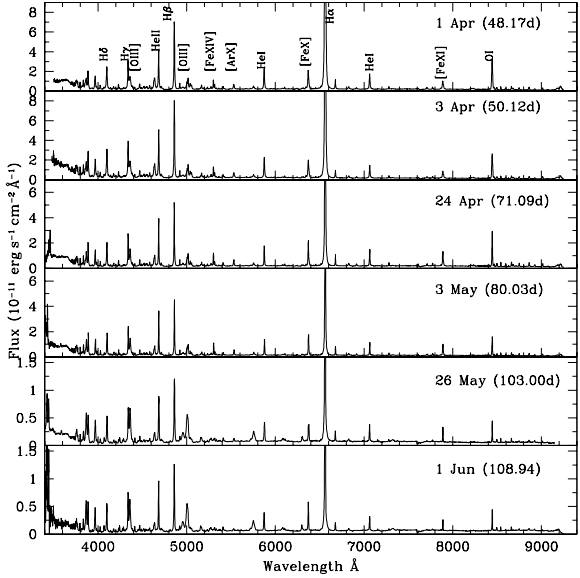}}
\caption[]{Spectra of RS Oph during 2006 April -- June.  The coronal and other high excitation lines are strong during this period.}
\label{specaprjun}
\end{figure}

\begin{figure}
\resizebox{\hsize}{!}{\includegraphics{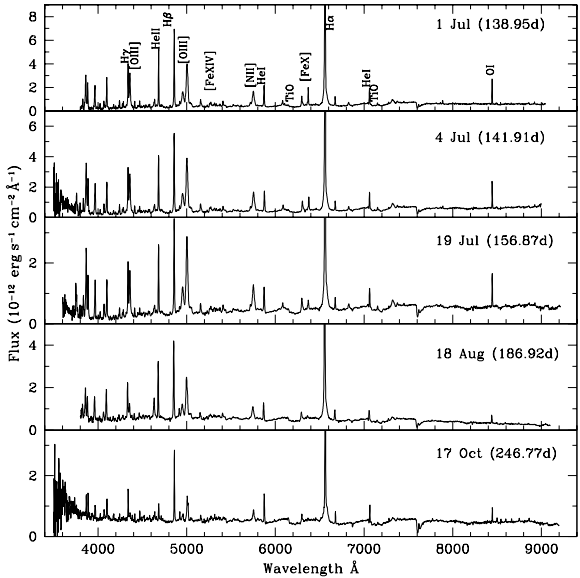}}
\caption[]{Spectra of RS Oph during 2006 July -- October. The nova has reached the nebular phase during this period. TiO absorption features are present.}
\label{specjuloct}
\end{figure}

\subsection{The nebular phase : beyond day 100}

The nebular phase began by day 100, and the nova was well into the nebular phase by $\sim 135$ days. The nebular lines are broader compared to the recombination lines. Coronal lines have weakened in strength. Other nebular lines such as [Ca II] 7323 begin to appear. Fe II and N I lines are absent or very weak. The
nebular lines get stronger between days 140--160 and are broader than other lines. H$\alpha$ shows asymmetric broad wings, with  a velocity similar to
those of the nebular lines. Contribution from the giant secondary is seen by day 140, as indicated by the presence of TiO absorption bands. These spectra
can be seen in Figure 6.

\section{Description of some emission lines}
\label{description}

\subsection{Emission line velocities}

The emission line widths are found to narrow with time. The hydrogen Balme lines indicate the
following velocities: 4200 km s$^{-1}$ (+1d); 3800 km $s^{-1}$ (+3d);
2000 km s$^{-1}$ (+5d); 1060 km s$^{-1}$ (+18d); 740 km s$^{-1}$ (+31d);
332 km s$^{-1}$ (+70d). Figure 7
shows the evolution of the emission line
velocity. The velocities vary as $t^{-0.4}$ for $t<5$ days and
$t^{-0.66}$ during $t=5-70$ days. This phenomenon is seen in the near-infrared lines as well and has been generally interpreted as free expansion of the shock front during the first four days and a deceleration phase thereafter (Das et al. 2006 and Banerjee et al.2009). These findings are also consistent with the X-ray observations (Bode et al. 2006, Sokoloski et al. 2006).

A slight increase in the emission line width of the recombination lines is seen after day 80. The nebular lines are broader than recombination lines, with a
velocity $v\sim 1500$ km s$^{-1}$ compared to $v\sim 400$ km s$^{-1}$ for the
recombination lines.

\begin{figure}
\resizebox{\hsize}{!}{\includegraphics{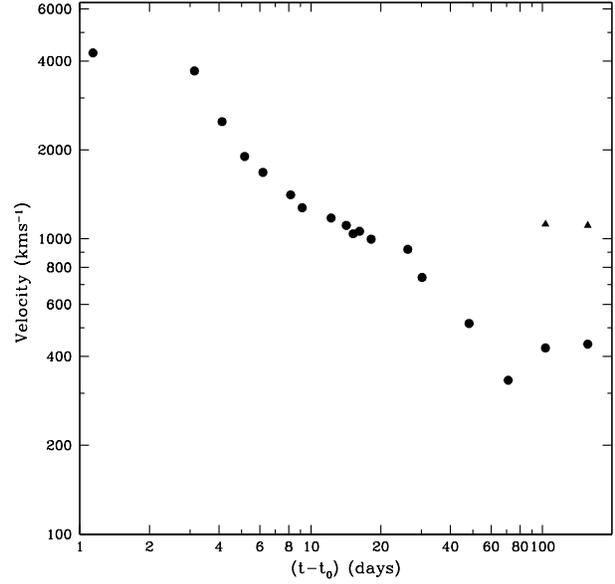}}
\caption[]{Velocities in the RS Oph spectra. It can be seen that the free expansion phase lasted only 4 days. The sharp fall in velocity after that is due to the shock wave generated when the high-velocity nova ejecta interact with the slow-moving red-giant wind. Note the increase in the
velocity after day 80.}
\label{vel}
\end{figure}

\subsection{Oxygen lines}

The O I 8446 \AA\, line is stronger than the 7774 \AA\, line on all days implying that Ly$\beta$ fluorescence is the dominant excitation mechanism. The evolution
of these two lines can be seen in Figures 8 and 9.
The [O I] 6300 and 6364 lines are best seen in spectra starting 2006 May. The O IV 6106
and 6182 lines are seen from 2006 March 10 onwards and are clearly identifiable till 2006 July. There is a hint of the [O III] 4363 line in the spectra obtained
from 2006 February 21 onwards. It is clearly seen on 2006 April 1. On 2006 May 26 it is seen to be
as strong as H$\gamma$. It faded thereafter, but is seen till 2006 August 18. There is no sign of this line in the 2006 October 17 spectrum. The evolution of
this line can be seen in Figure 10.
Similarly, there are indications of the  [O III] 4959 and 5007 lines  from 2006 February 20 onwards. They are
clearly seen on 2006 April 1, become extremely broad on 2006 May 26, are very prominent on 2006 July 4, and fade thereafter. The evolution of these lines can
be seen in Figure 11.

\begin{figure}
\resizebox{\hsize}{!}{\includegraphics{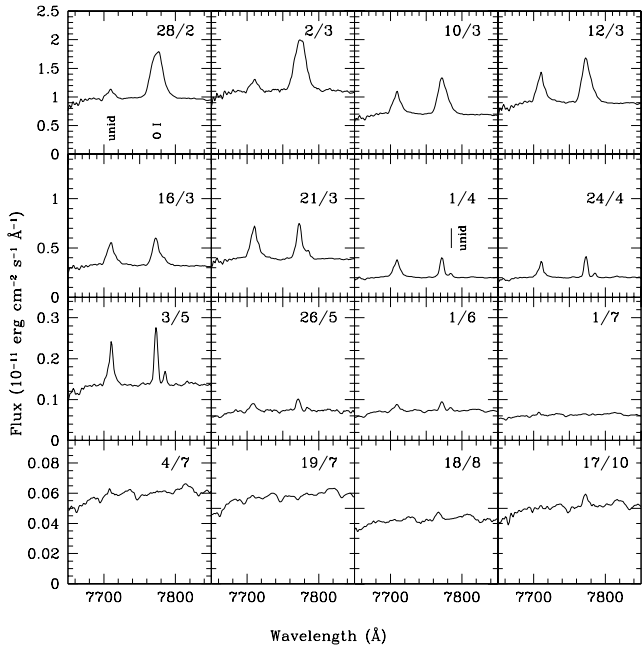}}
\caption[]{Evolution of the O I 7774 \AA\ line (top to bottom, left to right). 
Dates are in the dd/mm format. The stronger line in the first panel (top left) 
is O I. On its left is the unidentified line at 7710 \AA. The  development of 
another unidentified line at 7785 \AA\ can be seen (second panel from top). The 
O I line is seen to fade away in 2006 July, reappear in 2006 August and persist until 2006 October.}
\label{oi7774}
\end{figure}

\begin{figure}
\resizebox{\hsize}{!}{\includegraphics{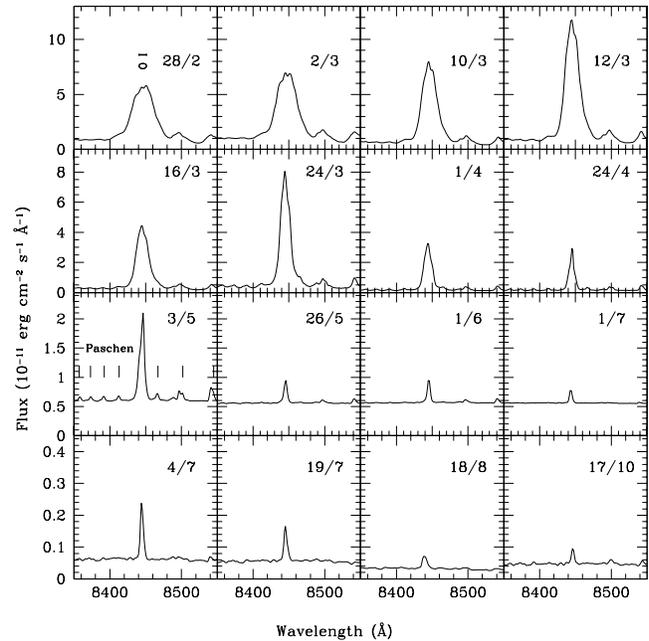}}
\caption[]{Evolution of the O I 8446 \AA\ line (top to bottom, left to right). 
Dates are in the dd/mm format. The stronger line in all the panels is the O I 
line. Paschen lines can also be seen; they are narrow and clearly seen on 2006 
May 3.}
\label{oi8446}
\end{figure}

\begin{figure}
\resizebox{\hsize}{!}{\includegraphics{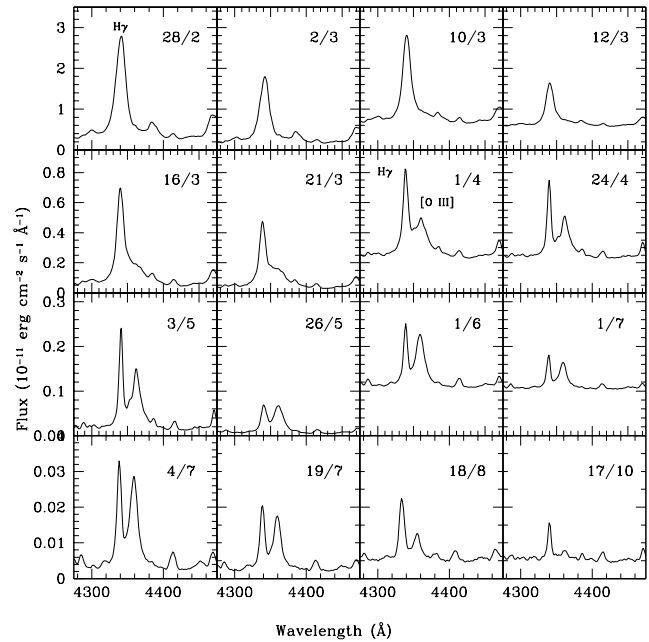}}
\caption[]{Evolution of the [O III] 4363 \AA\ line (top to bottom, left to 
right). Dates are in the dd/mm format. Spectra in the last two columns in the 
first three rows have been scaled up for clarity. The strong line in the first 
panel (top left) is H$\gamma$. The 4363 \AA\ line starts affecting the blue 
wing of H$\gamma$ from 2006 March 16, and is clearly seen on 2006 April 1; it 
has almost disappeared by 2006 October 17. }
\label{oiii4363}
\end{figure}

\begin{figure}
\resizebox{\hsize}{!}{\includegraphics{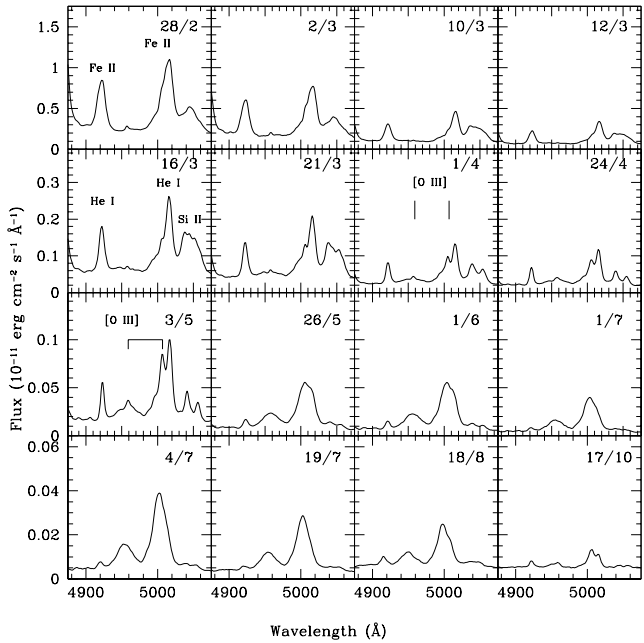}}
\caption[]{Evolution of the [O III] 4959, 5007 \AA\ lines (top to bottom, left 
to right). Dates are in the dd/mm format. Lines seen in the first panel (top 
left) belong to Fe II; lines belonging to other species at similar wavelengths 
strengthen with time. By 2006 March 16, He I (the first two lines) has 
strengthened; the third (broad) line is mostly Si II. The [O III] 5007 line can 
be seen feebly on the blueward side of the He I 5016 line. The [O III] 4959 and 
5007 lines can be clearly seen in the spectrum obtained on 2006 May 3; they 
persist until 2006 October 17.}
\label{oiii5007}
\end{figure}

\subsection{Helium lines}

There is only a hint of the He I 4471 \AA\ line on 2006 February 13; it shows 
up clearly two days later. The 5876 \AA\, line is very broad with a sharp central
absorption on 2006 February 13. It has contributions from some other line on its redward side till 2006 April. On 2006 February 13, the 6678 \AA\ line is very
broad with a central absorption, which persists  until early 2006 March. The 7065 \AA\ line shows a central absorption on 2006 February 13 which shifts to the
redward side on subsequent days. By 2006 March 16, the line has a Gaussian-like appearance. All the He I lines persist until 2006 October 17; the 7065 \AA\ line
shows a P-Cygni profile from 2006 July onwards (see Figure 12).
The He II 4686 \AA\ line appears on 2006 February 27 with a narrower FWHM than Balmer lines;  emission lines at this position in earlier spectra are likely to
be contributions from Fe II. By 2006 March 10, its width is similar to H$\beta$. This line persists until 2006 October 17 (see Figure 13).

\begin{figure}
\resizebox{\hsize}{!}{\includegraphics{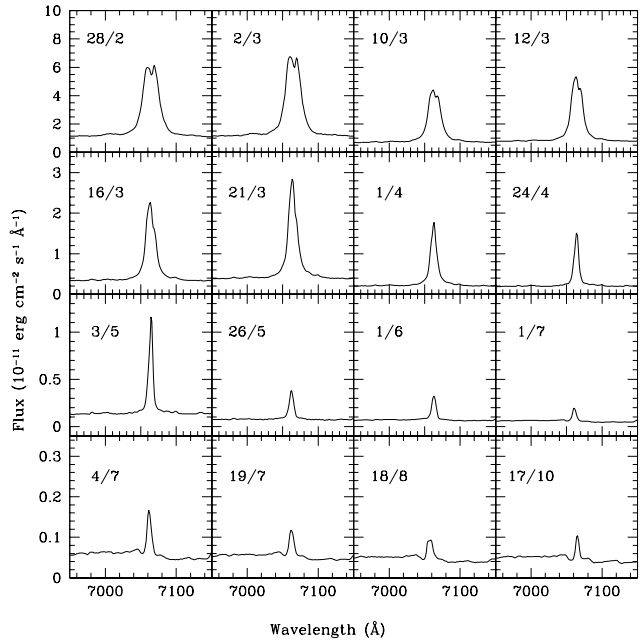}}
\caption[]{Evolution of the He I 7065 \AA\ line. Dates are in the dd/mm format.}
\label{hei7065}
\end{figure}

\begin{figure}
\resizebox{\hsize}{!}{\includegraphics{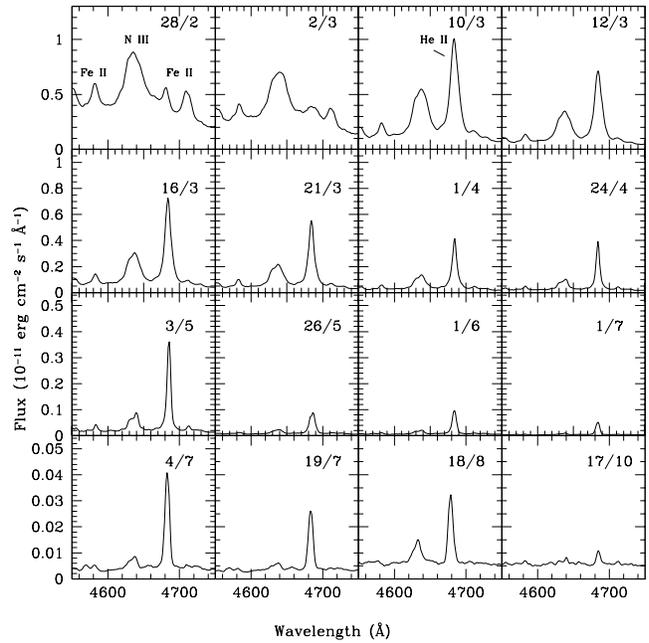}}
\caption[]{Evolution of the He II 4686 \AA\ line (top to bottom, left to right). Dates are in the dd/mm format. Lines seen in the first panel (top left) belong to Fe II and N III (the strongest line). The strong, narrow He II line is clearly seen by 2006 March 10 and persists till 2006 October 17. }
\label{heii4686}
\end{figure}

\subsection{Nitrogen lines}

The [N II] 5755 \AA\ line is present since 13 Feb. It is very broad initially 
but narrows down by 15 February. The 5660 \AA\ appears on 16 February and is 
broad, blended with the 5680 \AA\ feature. From 20 February onwards it has a narrow component on top of a broad 
feature. Other nitrogen lines show up after 26 February. The 5755 \AA\ line has 
a triangular shape and is seen to narrow with time. It suddenly broadens on 
26 May. This line persists until 2006 October 17 (See Figure 14).
The `4640' complex is observed from 2006 February to 2006 October (see Figure 13).

\begin{figure}
\resizebox{\hsize}{!}{\includegraphics{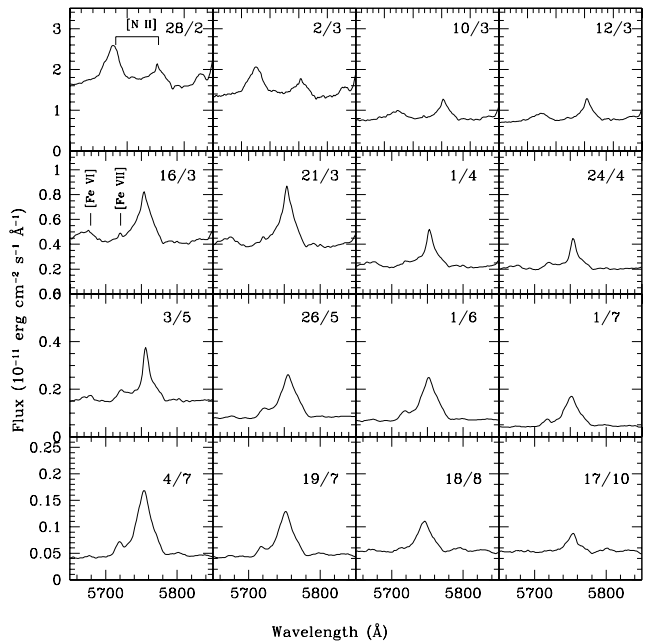}}
\caption[]{Evolution of the [N II] 5680 and 5755 \AA\, lines (top to bottom, left to right). Dates are in the dd/mm format. Both the lines (5660~+~5680 blend and 5755) are seen in the first panel. The 5660 / 5680 line weakens and is replaced by the [Fe VI] 5679 line in later spectra. The [Fe VII] 5721 line shows as a small kink on the redward side of 5755 on 2006 March 16, become stronger and fades away by 2006 August 18.}
\label{nii5755}
\end{figure}

\subsection{Coronal lines}

Coronal lines in RS Oph were first seen and identified by Adams \& Joy (1933) on day 51 during the 1933 outburst. The presence of five coronal lines at 3987, 4086, 4231, 5303 and 6374 \AA\ could be well-established on their spectrograms obtained on later nights. In this outburst, coronal lines of neon ([Ne III], [Ne V]), argon ([Ar IV], [Ar X], [Ar XI], [Ar XIV]) and iron ([Fe VII], [Fe X], [Fe XI], [Fe XIV]) can be clearly seen in the
spectra obtained during 2006 March -- June. It appears that the calcium lines showed an unusual behaviour during this outburst as compared to the previous
outbursts. [Ca XIII] 4086 \AA\ was absent, [Ca XV] 5694 \AA\ was not prominent. The temporal development of the iron  green and red coronal lines are
shown in Figures 15 and 16 respectively.
Many coronal lines are close in wavelength to other permitted or low ionisation lines -- for example, [Ne III] 3970 \AA\ and H$\epsilon$, [Ni XII] 4231 \AA\ and Fe II, [Ar XIV] 4412 and [Fe II], etc -- and it becomes difficult to separate the contributions from the individual components from observations of emission lines.  However, it is very likely that the permitted or low ionisation lines are the dominant contributors in the early outburst spectra.

\begin{figure}
\resizebox{\hsize}{!}{\includegraphics{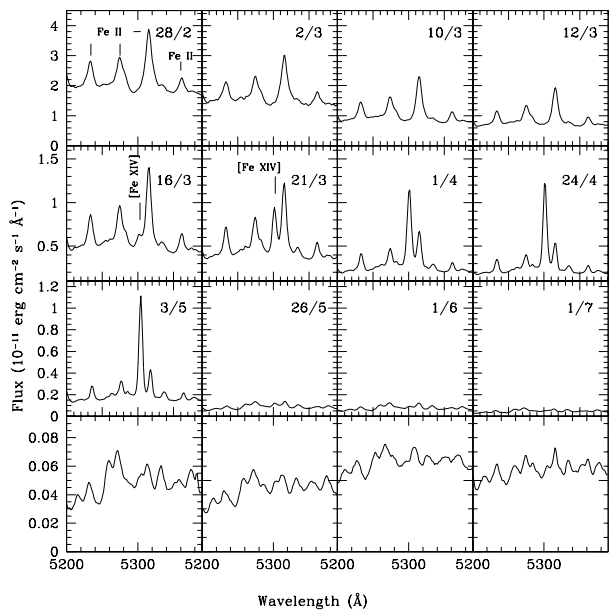}}
\caption[]{Evolution of the [Fe XIV] 5303 \AA\ line (top to bottom, left to right). Dates are in the dd/mm format. Fe II lines are seen in the first panel (top left). The [Fe XIV] appears as a small kink on the redward side of the Fe II 5317 line on 2006 March 16, increasing in strength by 2006 May 3, and barely seen by 2006 June 1. The last row of spectra display just the fluctuations at the continuum level.}
\label{fe5303}
\end{figure}

\begin{figure}
\resizebox{\hsize}{!}{\includegraphics{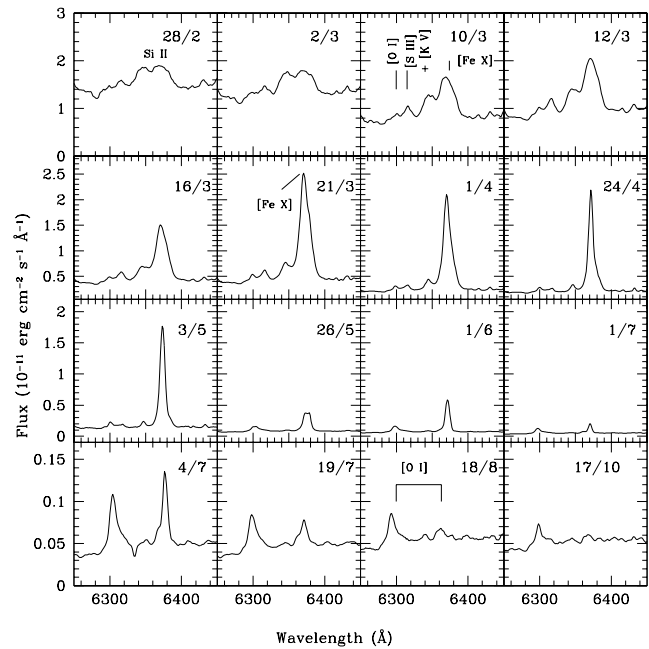}}
\caption[]{Evolution of the [Fe X] 6374 \AA\ line (top to bottom, left to right). Dates are in the dd/mm format. The broad lines seen in the first panel (top left) are mostly Si II. The [Fe X] 6374, [S III] + [K V] 6315 are seen on 2006 March 10. The [O I] 6300 line is also feebly seen on this day. By 2006 August 18, only the [O I] 6300 and 6363 lines are clearly seen.}
\label{fe6374}
\end{figure}

\subsection{The Raman scattered 6830, 7088 \AA\ lines}
\label{raman-scatt}

The strong line at 6830 \AA\ which was first seen in the 1933 outburst was attributed to [Kr III] (Joy \& Swings 1945). This line was seen in other outbursts also. We now know that this emission feature, seen in many symbiotic stars, is due to the Raman scattering of the O VI 1032 \AA\ line by neutral hydrogen (Schmid 1989). Schmid also noted that in some symbiotic stars a fainter companion feature appears at 7088 \AA. This is due to the same process as for the 1038 \AA\ line of the O VI resonance doublet. Schmid has suggested that this Raman scattering in symbiotic stars takes place in the atmosphere of the cool giant and in the inner parts of its stellar wind. The presence of the Raman scattered O VI lines implies the existence of a hot ionising source, the WD on which hydrogen is still burning after the outburst.
The 2006 outburst is the first one since this identification was done.
This line is first seen in our  spectrum of 2006 March 10 and persists until 2006 August 18. In our next spectrum obtained on 2006 October 17, this
line is probably not seen; other nearby lines dominate this wavelength region (see Figure 17).
The 7088 line is also feebly seen in the wings of He I 7065 \AA\ during the same period.

\begin{figure}
\resizebox{\hsize}{!}{\includegraphics{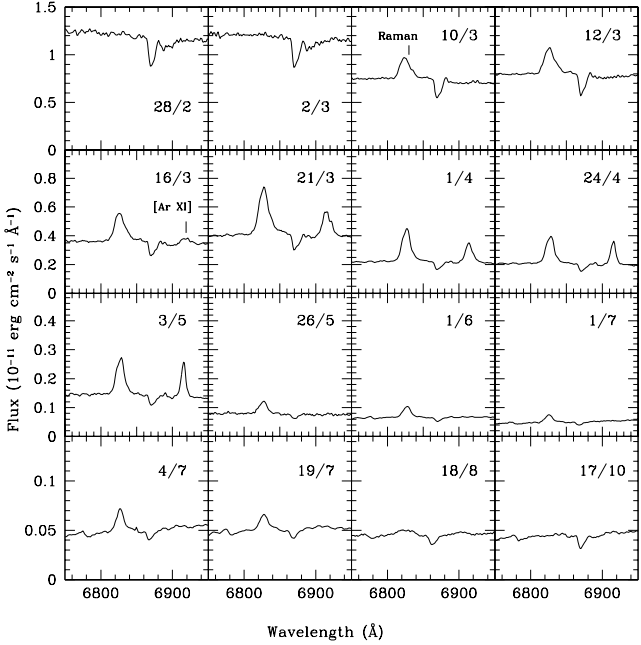}}
\caption[]{Evolution of the Raman scattered 6830 \AA\ line (top to bottom, left to right). Dates are in the dd/mm format. This line appears suddenly on 2006 March 10; there is no gradual strengthening as seen for other lines. It disappears after 2006 July 19. The [Ar XI] 6919 line is seen during 2006 March 16 to 2006 May 3.}
\label{raman}
\end{figure}

\subsection{Formation of the spectral lines}

{\it Swift} observations have revealed the existence of a supersoft X-ray phase during days 26--90 of the outburst  which is due to the matter 
burning quiescently on the WD (Osborne et. al. 2011). This source peaks at around day 50, similar to the behaviour of the coronal 
lines. This indicates that some of the flux in the coronal lines, especially during days 26--90, must be due to photoionisation. As mentioned in 
Section 4.6, the behaviour of the Raman lines also corroborate this idea. The He II and O IV lines having high ionisation 
potentials also were strong during 2006 March -- June. Similarly, multi-frequency radio data show that the spectral index is consistent with a 
mixed non-thermal and thermal emission during days 5--60 (Eyres et. al., 2009). Therefore, as for other novae, optical spectra should show 
the characteristics of photoionised ejecta. Photoionisation could even be the dominant mechanism at least during days 29--90. So, in Section 5, we attempt the photoionisation modelling of our optical spectra using CLOUDY.

\section{Photoionisation model analysis}
\label{cloudy-analysis}

We use the CLOUDY photoionisation code,  c13.05 (Ferland et al., 2013) to model the emission line spectra of RS Oph observed during and 
after its recent outburst in 2006. We have chosen representative spectra on 9 epochs covering a duration of $\sim$ 438 days. Modelling of 
data over such a long time period enables sampling over a broader range of ionisation and excitation levels in the emission lines, and thus 
helps to constrain the results more accurately. This method has been used previously to determine elemental analysis and physical 
characteristics of a few novae by modelling the observed spectra, for example, QU Vul (Schwarz et al. 2002), V1974 Cyg (Vanlandingham et 
al. 2005), V838 Her $\&$ V4160 Sgr (Schwarz et al. 2007), V1065 Cen (Helton et al. 2010), RS Oph (Das \& Mondal, 2015).
CLOUDY generates model spectra assuming a non-LTE, spherically expanding ejecta illuminated by a central source, by solving the 
equations of thermal and statistical equilibrium, using a set of input parameters. The calculations incorporate effects of important ionisation 
processes, 
e.g., photo, Auger, collisional, and recombination process viz. radiative, dielectronic, three-body recombination, and charge transfer. The 
parameters specify the initial physical conditions of the source and the ejected shell; the source parameters are the spectral energy 
distribution of the continuum source which is assumed to be a black body, its temperature and luminosity; the shell parameters are density, 
inner and outer radii, geometry, covering factor (fraction of 4$\pi$ sr enclosed by the model shell), filling factor (ratio of the contribution of 
dense shell to diffuse shell) and elemental abundances (relative to solar). The shell density $n(r)$ which is set by hydrogen density, and the 
filling factor $f(r)$ may vary with radius, as given by the following relations,\\
\begin{equation}
n(r) = n(r_{0}) (r/r_{0})^{\alpha} cm^{-3} \,\,\,\,\, \,\,\&\,\,\,\,\,\, f(r) = f(r_{0}) (r/r_{0})^{\beta},
\end{equation}

where, $r_{0}$ is the inner radius, $\alpha$ and $\beta$ are exponents of power laws.
We choose $\alpha = -3$, the filling factor $= 0.1$ and filling factor power-law exponent, ($\beta$) = 0, which
are the typical values used in other CLOUDY studies for novae. 
The inner ($R_{\rm{in}}$) and outer ($R_{\rm{out}}$) radii of the ejected spherical shell are held constant to reduce the number of free 
parameters. $R_{\rm{in}}$ and $R_{\rm{out}}$ on each epoch were calculated using the values of minimum ($V_{\rm{min}}$) and maximum 
($V_{\rm{max}}$) expansion velocities, respectively.
We adopt initial shell velocities as $V_{\rm{min}} = 3500$~km s$^{-1}$ and $V_{max} = 4500$~km s$^{-1}$ during explosion, based on the 
optical and near-infrared
observations (this work, Das et al. 2006 and Skopal et al. 2008). The velocities decrease with time as $t^{-0.40}$ for $t \le 5$ days, and as 
$t^{-0.66}$ for $t = 5 - 70$ days. Using these relations, we calculate the minimum and maximum velocities at each epoch, which are then used
to calculate the values of $R_{\rm{in}} $ and $R_{\rm{out}}$ at each epoch. We assume the continuum shape to be a blackbody of high 
temperature, $10^{5.95} \ge T_{\rm{BB}} \ge 10^{4.2}$ K, as assumed in the previous investigations, to ensure that it supplies the correct 
amount of photons for photoionisation. Based on parameter space discussed in the above, the CLOUDY output predicts the model fluxes of 
different emission lines, which are compared with the observed fluxes of emission lines after normalizing with respect to H$\beta$.

Several spectra are generated by varying the free parameters, e.g. hydrogen density, underlying luminosity, effective blackbody temperature,
and abundances of only those elements with observed lines. The abundances of all other elements which do not show lines in the observed 
spectra were fixed at solar values. Due to the inhomogeneous nature of novae ejecta, we assume a two-component model in the initial 
outburst phase: a higher density shell to best fit the lower ionisation lines, and a lower density shell to fit the higher ionisation lines.
To reduce the number of free parameters, we use the same parameters for each component, except
hydrogen densities at the inner radius and the covering factors, assuming that the sum of the two covering factors be less than or equal to 1. 
This increases the number of total free parameters by 2 due to the second component's initial density and covering factor.
The final model line ratios were calculated by adding each component's line ratio after multiplying by its covering factor.
Fluxes of the model generated lines were then compared with fluxes of the observed lines and the best fit model were chosen by calculating 
$\chi^{2}$ and reduced $\chi^{2}$ ($\chi^{2}_{\rm{red}}$) of the model given by,
\begin{equation}
\chi ^{2} = \sum\limits_{i=1}^n (M_{i} - O_{i})^{2}/ \sigma_{i}^{2}, \,\,\,\,\&\,\,\,\, \chi^{2}_{\rm{red}} = \chi^{2}/\nu
\end{equation}
where, $n$ is the number of observed lines, $n_{p}$ is the number of free parameters, $\nu$, the degree of
freedom given by $n - n_{p}$, and $M_{i}$ is the modelled ratio of line flux to hydrogen line flux, $O_{i}$ is the measured flux ratio,
and $\sigma_{i}$ is the error in the observed flux ratio. We estimate error in the range of $10 - 30\%$, depending upon the intensity of a
spectral line, possibility of blending with adjacent lines,
and error in the measuring line fluxes. Values of $\chi^{2}$ $\sim$ $\nu$ and
($\chi^{2}_{\rm{red}}$) value should be low (typically in the range of 1 - 2) for a good fit
model.

\section{Results and Discussions}

\subsection{Early Phase: 2006 February -- May}

Relative fluxes of the best-fit model predicted lines, observed lines and 
corresponding $\chi^{2}$ values during the early phase are presented in Table 2.
Here, we have considered the lines which are present both in the model generated and observed spectra for the calculation of $\chi^{2}$ 
values. The observed line fluxes have been determined using IRAF tasks; the profiles which have multiple 
components have been decomposed with multiple Gaussians using IRAF tasks.
To minimize errors associated with flux calibration between different epochs, the modelled and observed flux ratios have been calculated 
relative to H$\beta$. The values of best-fit parameters are presented in Table 4. The best fit modelled spectra (grey lines) with  the 
observed optical spectra  (black lines) during this phase are shown in Figures 18 - 22.
The spectra, in the early phase, are dominated by prominent features of low ionisation lines, e.g. H$\epsilon$ (3970 \AA), H$\delta$ 
(4101 \AA), H$\gamma$ (4340 \AA), He II (4686 \AA), H$\beta$ (4863 \AA). There are also some higher ionisation lines, e.g. [Fe VII] 
(3760 \AA), Fe II and [Ni  XII] (4233 \AA), He I and [Ar IV] (4713 \AA), [Ar X] (5535 \AA), [Fe XI]  (7890 \AA), [Fe X] (6374 \AA), [Fe XIV] 
(5303 \AA) etc., which started to appear at the end of the early phase. Since, the gas is highly 
inhomogeneous at this time, a one component model was unable to fit the H$\alpha$/H$\beta$ flux ratio properly. So, we had to consider a   
two component model. The first component (clump) consisted of a highly dense cloud to fit the lower ionisation lines. This component was 
optically thick, and Case-B was taken into consideration as well. Majority of the observed lines were generated using this component, but it 
under-represents the highest ionisation lines. The second component (diffuse), with a lower density, was less ($10 - 30\%$) in volume, but it 
allowed the source photon to ionise some of the species to their higher ionisation states (e.g. [Ar IV] (4713 \AA), [Ar X] (5535 \AA)), while not
changing the clump contribution to the spectra.

Initially, in 2006 February, the ionising source is surrounded by a dense ($n_H \sim 10^{11}$ cm$^{-3}$) cloud. So, the effective temperature 
and luminosity ($T \sim 10^{4.45}$ K, $L \sim 10^{36.8}$ erg s$^{-1}$) of the source is relatively lower. As the ejecta expands, more energy 
from the central engine comes out. Therefore, while the ejecta density decreases ($n_H \sim 10^{8.41}$ cm$^{-3}$), the blackbody 
temperature ($T_{BB}$) and the luminosity increases to $T_{BB} \sim 10^{5.6}$ K and L $\sim 10^{37.1}$ erg s$^{-1}$, respectively in 2006 
May to photoionise more matter. The estimated elemental abundances during this phase are given as logarithm of the numbers relative to 
hydrogen and relative to solar in Table 4. The values show that helium, nitrogen, oxygen, neon, argon, iron and nickel 
abundances are all enhanced relative to solar whereas abundances of calcium and sulphur are at solar values. However, we must be cautious 
while considering the abundances of calcium, because calcium abundance was calculated using only the Ca V (6086 \AA) line observed on 
2006 May 26. This line is blended with [Fe VII]. So, first iron abundance was set using the other prominent Fe II (4180 \AA, 4415 \AA\ and 
4584 \AA) lines. Then, the calcium abundance was changed such that the total flux of the blended feature in the model matched the 
observed spectra.  Similarly, nitrogen abundance on day 28 was determined using the N II (5535 \AA) line blended with [Ar X]. But, on days 
50 and 103, there are two other N II (4242 \AA\ and 5755 \AA) lines that helped to set the nitrogen abundance correctly. These values match 
well with the results obtained by Das \& Mondal (2015).

\begin{table*}
\label{cloudyFebMay}
\setlength{\tabcolsep}{2pt}
\begin{center}
\caption[]{Observed and best-fit CLOUDY model line fluxes.}
\begin{tabular}{c c c c c c c c cccccccccccccccc}
\hline\noalign{\smallskip}
$\lambda$ &	      &          &	Feb. 21 &	         &	 	&  Feb. 24 &            &       &	Mar. 12  &	 	      &       &  Apr. 3     & & & May. 26  &  \\
 (\AA)  & Line  & Obs.     &  Mod.    & $\chi^{2}$ & Obs. &  Mod.     & $\chi^{2}$ & Obs.  &  Mod.     & $\chi^{2}$ & Obs.  & Mod.       & $\chi^{2}$ & Obs. & Mod.      & $\chi^{2}$\\
\hline\noalign{\smallskip}
3750    & H12     				& 0.04 & 0.04&0.00 & 0.02 & 0.05	&0.09  & ...  &	...	&...	& ...  &...	&...& ...  &...&...\\
3760    & [Fe VII]				& ...  & ...& ...& ...  &... & ...&	...&	...	&...	& 0.07 &	0.19	&1.62& 0.31 &0.14&2.85	\\
3771    & H11     				& 0.06 & 0.01&0.22& 0.03 &0.04&0.00&   ...&	...	&...	&...	   &...	&...& 0.03 &0.05&0.06\\
3798    & H10     				& 0.10 &	0.06	&0.18& 0.06	&0.06&0.00& 0.03 &	0.07	&0.17& 0.07 &0.06&0.02& 0.08 &0.10&0.08\\
3835    & H$\eta$ 				& 0.18 &	0.07	&1.17& 0.14 	&0.17&0.08 & 0.04 &	0.12	&0.61& 0.10 &0.14&0.21& 0.13 &0.13&0.0\\
3868    & [Ne III]				& ...  &	 ...	& ...& ...  & ...& ...& ...&	...	&...	& 0.20 &0.14	&0.42& 0.62 &0.49&1.67\\
3889    & H$\zeta$, He I 		& 0.27 &	 0.14&1.82& 0.22	& 0.20&0.06	& 0.12 &0.15	&0.10& 0.26 &0.29&0.10& 0.38 &0.29&0.7\\
3970    & H$\epsilon$, [Ne III]  & 0.26 &0.21&0.29 & 0.18	& 0.13&0.26 	& 0.12 &	0.16	&0.30& 0.26 &0.19&0.46& 0.36 &0.21&2.22\\
4026    & He I, He II  			& 0.05 &0.03	&0.04& 0.05	& 0.03&0.06 	& 0.04 &	0.06	&0.04& 0.05 &0.06&0.02& 0.05 &0.06&1.83\\
4070    & [S II]                 & ...  & ...& ...& ...	& ...&...&...  &... &... & ...  &... &...& 0.10 &0.03&0.55\\
4101    & H$\delta$   			& 0.34 &	0.32	&0.02& 0.28	&0.29&0.02 	& 0.22 &	0.19	&0.09& 0.33 &0.33&0.0& 0.05 &0.28&0.48\\
4144		& He I					& ...  & ...& ...& ...	& ...&...&...  &... &... &...  &...  &...& 0.03 &0.02&0.02\\
4180    & Fe II, [Fe II] 		& ...  &	 ...	& ...& ...  & ...& ...& ...&	...	&...	& 0.05 &	0.02	&0.10& 0.06 &0.03&0.09\\
4233    & Fe II, [Ni XII] 		& 0.05 &	0.06&0.00& 0.03	& 0.04&0.00 	& 0.03 &0.03	&0.00& 0.06 &0.04&0.03& 0.07 &0.05&0.04\\
4242    & N II  					& ...  &	 ...	& ...& ...	& ...& ...& ...&	...	&...	& 0.02 &	0.04	&0.05& 0.09 &0.07&0.05\\
4340    & H$\gamma$       		& 0.45 &	0.22	&5.33& 0.39	&0.33&0.29	& 0.30 &0.34	&0.19& 0.34 &0.29&0.22& 0.52 &0.40&1.38\\
4415    & [Fe II] 				& ...  &	 ...	& ...& ...	& ...& ...& ...&...	&...	& 0.04 &	0.04	&0.00& 0.06 &0.07&0.00\\
4471    & He I 	 				& 0.11 &	0.07	&0.12& 0.09	&0.06&0.10& 0.05 &	0.11	&0.44& 0.06 &0.09&0.14& 0.11 &0.12&0.00\\
4584    & Fe II					& ...  &	 ...	& ...& ... &...&...& 0.03 &	0.03	&0.00& 0.04 &0.05&0.02& ... &...&...\\
4686    & He II 					& ... & ...& ...&... & ...& ...& 0.34 &	0.19	&2.22& 0.52 &0.37&1.9& 0.65 &0.52&1.82\\
4713    & He I, [Ar IV]			& ...  &	 ...	& ...& 0.04	& 0.03&0.00& 0.02 &	0.01	&0.03& 0.21 &0.07&1.9& ... &...&...\\
4863		& H$\beta$				& 1.00 & 1.00& 0.00& 1.00& 1.00& 0.00& 1.00& 1.00& 0.00& 1.00& 1.00& 0.00& 1.00& 1.00& 0.00\\
4922    & Fe II, He I			& 0.09 & 0.05&0.19&	0.09 &0.06&0.06& 0.07 &	0.06	&0.01& 0.06 &0.04&0.05& 0.09 &0.11&0.04\\
4959    & [O III] 				& ...  &	 ...	& ...&	...  &...&...& ... &	...	&...	& 0.06 &	0.05	&0.00& 0.46 &0.31&2.15\\
5007    & [O III] 				& ...  &	 ...	& ...&	...  &...&...& ... &	...	&...	& 0.23 &	0.14	&0.84& 1.06 &0.01&0.23\\
5016    & Fe II, He I   			& 0.15 &0.08	&0.43& 	0.18 &0.07 	&1.16 	&  ... &	...	&...	& ...  &...	&...& ...  &...&...\\
5169    & Fe II         			& 0.08 &	0.13	&0.28 &	0.09	 &0.08	&0.01	&  ... &	...	&...	& ...  &...	&...& ... &... &...\\
5235    & Fe II         			& 0.02 &	0.18	&2.57&	0.02	 &0.01	&0.01	&  ... &	...	&...	& ...  &...	&...& ... &... &...\\
5276    & Fe II         			& 0.03 &	0.08	&0.29 &	0.03	 &0.06	&0.08	&  ... &	...	&...	& ...  & ...	&...& ... &... &...\\
5303    & [Fe XIV]     			& ...  &	 ...	& ...&	...	 &...&...& ... &	...	&...	& 0.10 &	0.09	&0.02& ... &... &...\\
5317    & Fe II       			& 0.05 &	 0.04& 0.02 &	0.04	 & 0.03  &0.00	&  ... &	...	&...	& ...  & ...	&...& ... &... &...\\
5412    & He II 					& ...  &	 ...	& ...&	...	 &...&...&  ... &...	&...	& 0.06 &	0.10	&0.19& 0.09 &0.09&0.00\\
5535    & N II, [Ar X] 			& ...  &	 ...	& ...&	... 	 &...&...& 0.02 &0.02&0.00& 0.09 &0.02&0.47	& 0.08 &0.04&0.15\\
5755    & [N II] 				& ...  &	 ...	& ...&	... 	 &...&...& ...  &...	&...	& 0.07 &	0.04	&0.12& 0.43 &0.32&1.26\\
5876    & He I        			& 0.31 &	0.30	&0.00 & 	0.44	 &0.24	& 3.98& 0.50 &	0.46	&0.16& 0.29 &0.34&0.28& 0.29 &0.43&.89\\
6086    & [Fe VII], [Ca V] 		& ...  &	 ...	& ...&	...  &...&...& ...	&...	&...	& ...  & ...	&...& 0.20 &0.32&1.55\\
6101    & O IV					& ...  &	 ...	& ...&	... 	 &...&...& 0.03 &0.03&0.00& ... &...&...& ... & ...&...\\
6300		& [O I]					& ...  & ...& ...& ...	& ...&...&...  &... &...& ...  & ...&... & 0.12 &0.18&0.37\\
6374    & [Fe X] 				& ...  &	 ...	& ...&	...  &...&...& ...  &...	&...	& 0.32 &	0.07	&6.27& 0.33 &0.29&0.16\\
6563    & H$\alpha$   			& 4.74 &	4.45	&8.26 & 	6.12	 & 5.76	&12.53& 9.60 &	9.24	&12.72& 5.81 &6.15&12.08	& 4.46 &4.75&8.83\\
6678    & He I        			& 0.09 & 0.08&0.02& 	0.14	 & 0.06	&0.64& 0.17 &	0.12	&0.23& 0.08 &0.1	&4.62& 0.12 &0.12&0.00\\
6830    & Raman 					& ...  &	 ...	& ...&	...	 &...&...& 0.03 &0.01&0.03& ... &...&...& ... &...&...\\
7065    & He I		  			& 0.14 & 0.16&0.02& 	0.24 & 0.16 	&0.63& 0.37 &	0.35	&0.03& 0.19 &0.27&0.65& 0.21 &0.37&0.67\\
7280    & He I        			& 0.03 & 0.08&0.23& 	0.07	 & 0.06	&0.00& 0.03 &	0.04	&0.01& 0.02 &0.04&0.05& 0.02 &0.05&0.13\\
7890    & [Fe XI] 				& ...  & ...	& ...&	...	 &...&...&... &...&...	& 0.15 &	0.02&1.65& 0.16 &0.06&0.96\\
8598    & Pa 14       			& 0.03 & 0.04&0.00 & 	0.04	 &0.03	&0.00	& 0.05 &	0.03	&0.05& 0.02 &0.02&0.0& 0.02 &0.03&0.02\\
8750    & Pa 12        			& 0.04 & 0.03&0.00 & 	0.06	 & 0.06	&  0.00 & 0.07 &0.04&0.07& 0.03 &0.03&0.0& 0.02 &0.03&0.02\\
8862    & Pa 11 					& 0.06 &	0.04	&0.03 &	0.08	 &0.04 	&0.13   & 0.09 &	0.06&0.09& 0.03 &0.03&0.0& 0.02 &0.03&0.03\\
9015    & Pa 10 					& 0.09 & 0.10&0.02	 &	0.09 &0.11 	&0.03 	& 0.12 &	0.09	&0.09& 0.03 &0.06&0.09	& 0.01&0.07&0.42\\

\hline
Total &							&... & ...& 21.60  &... 	&... &  20.28  &...	&... &17.71	&...	&...	&	30.49&...	&...	&32.94\\
\noalign{\smallskip}\hline
\end{tabular}
\end{center}
\end{table*}

\begin{figure}
\label{cloudy21Feb}
\centering
\includegraphics[width= 3.3 in, height = 2.65 in, clip]{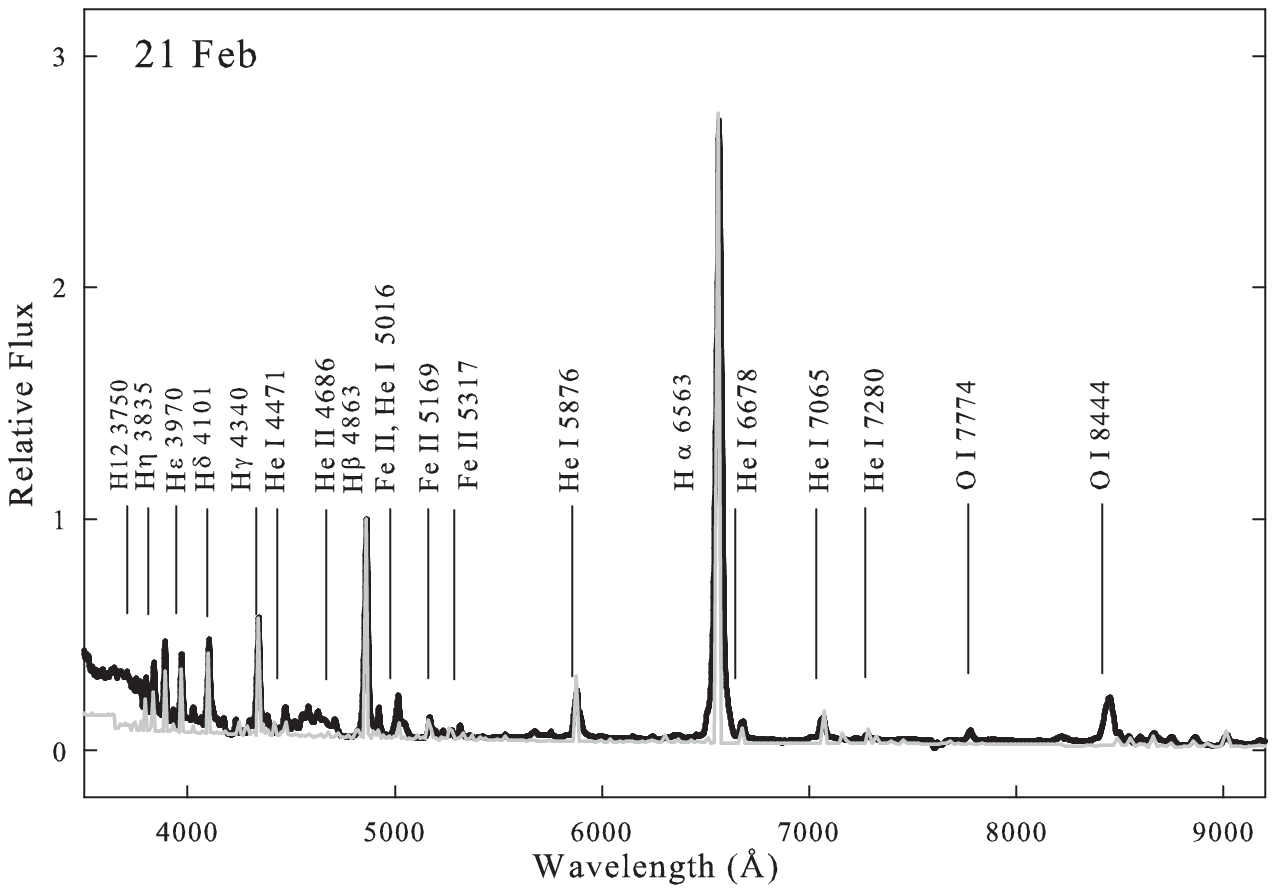}
\caption{Best-fit CLOUDY model spectra (grey line) plotted over the observed spectra (black lines) of RS Oph observed on 2006 February 21.
The spectra were normalized to H$\beta$. Also few of the strong features have been marked (see text for details).}
\end{figure}

\begin{figure}
\label{cloudy24Feb}
\centering
\includegraphics[width= 3.3 in, height = 2.65 in, clip]{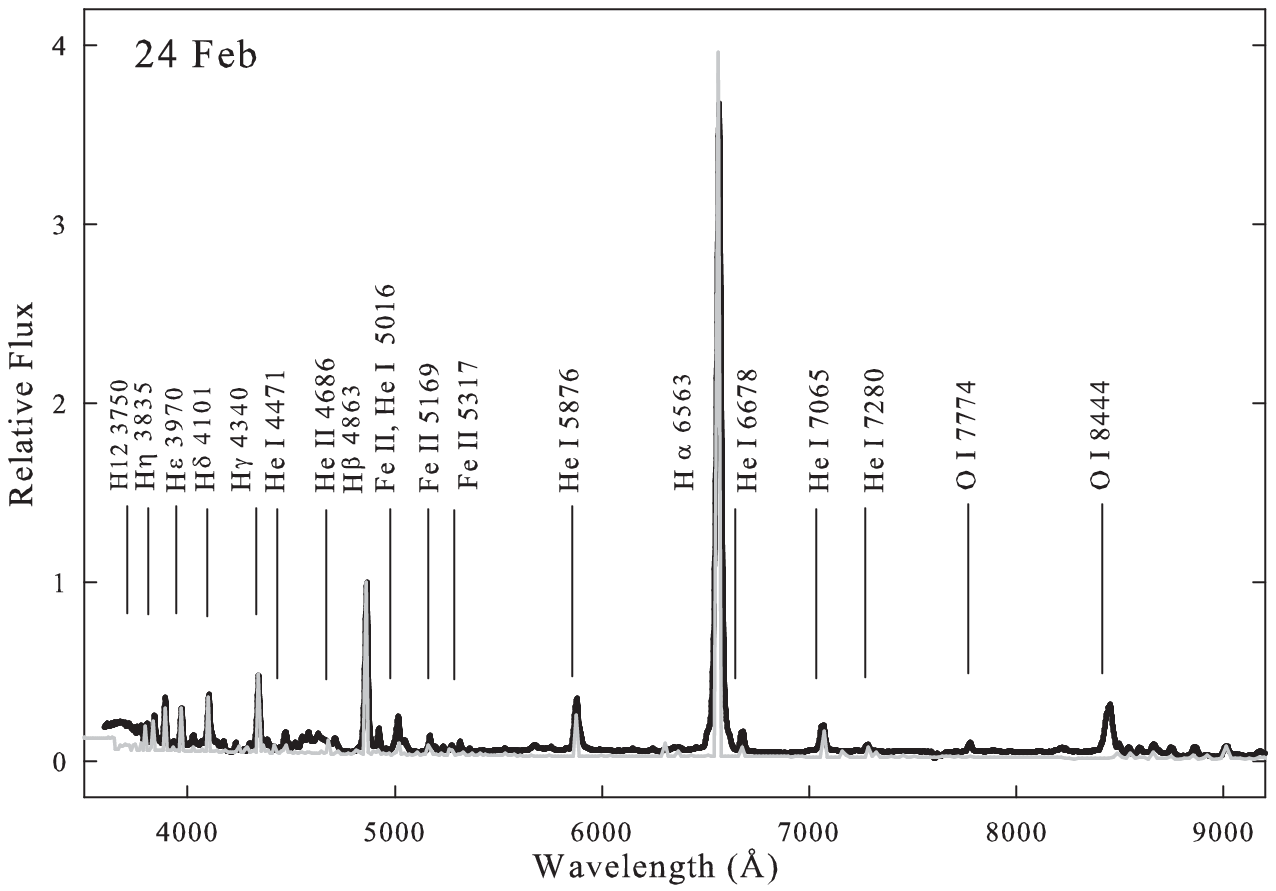}
\caption{Best-fit CLOUDY model spectra (grey line) plotted over the observed spectra (black lines) of RS Oph observed on 2006 February 24.
The spectra were normalized to H$\beta$. Also a few of the strong features have been marked (see text for details).}
\end{figure}

\begin{figure}
\label{cloudy12Mar}
\centering
\includegraphics[width= 3.3 in, height = 2.65 in, clip]{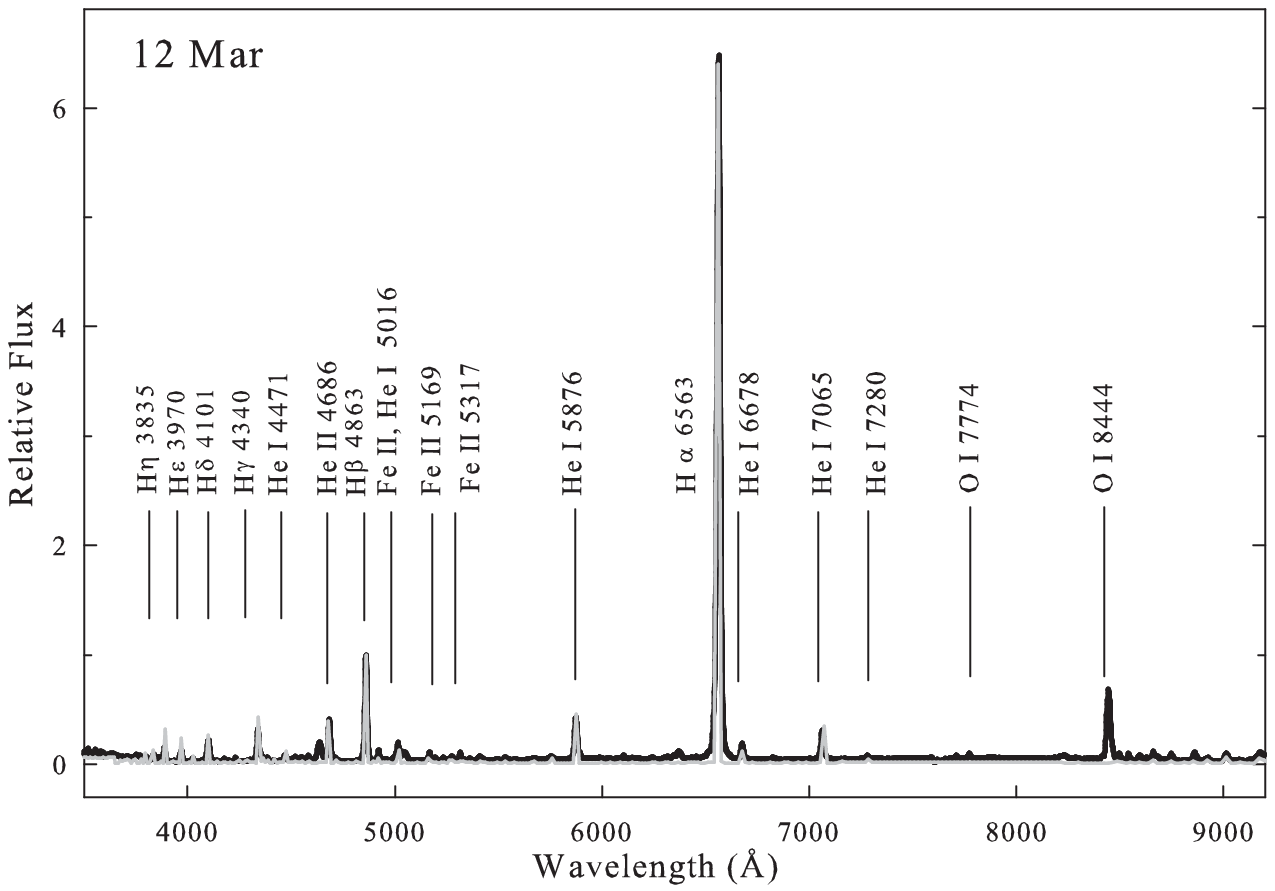}
\caption{Best-fit CLOUDY model spectra (grey line) plotted over the observed spectra (black lines) of RS Oph observed on 2006 March 12.
The spectra were normalized to H$\beta$. Also a few of the strong features have been marked (see text for details).}
\end{figure}

\begin{figure}
\label{cloudy3Apr}
\centering
\includegraphics[width= 3.3 in, height = 2.65 in, clip]{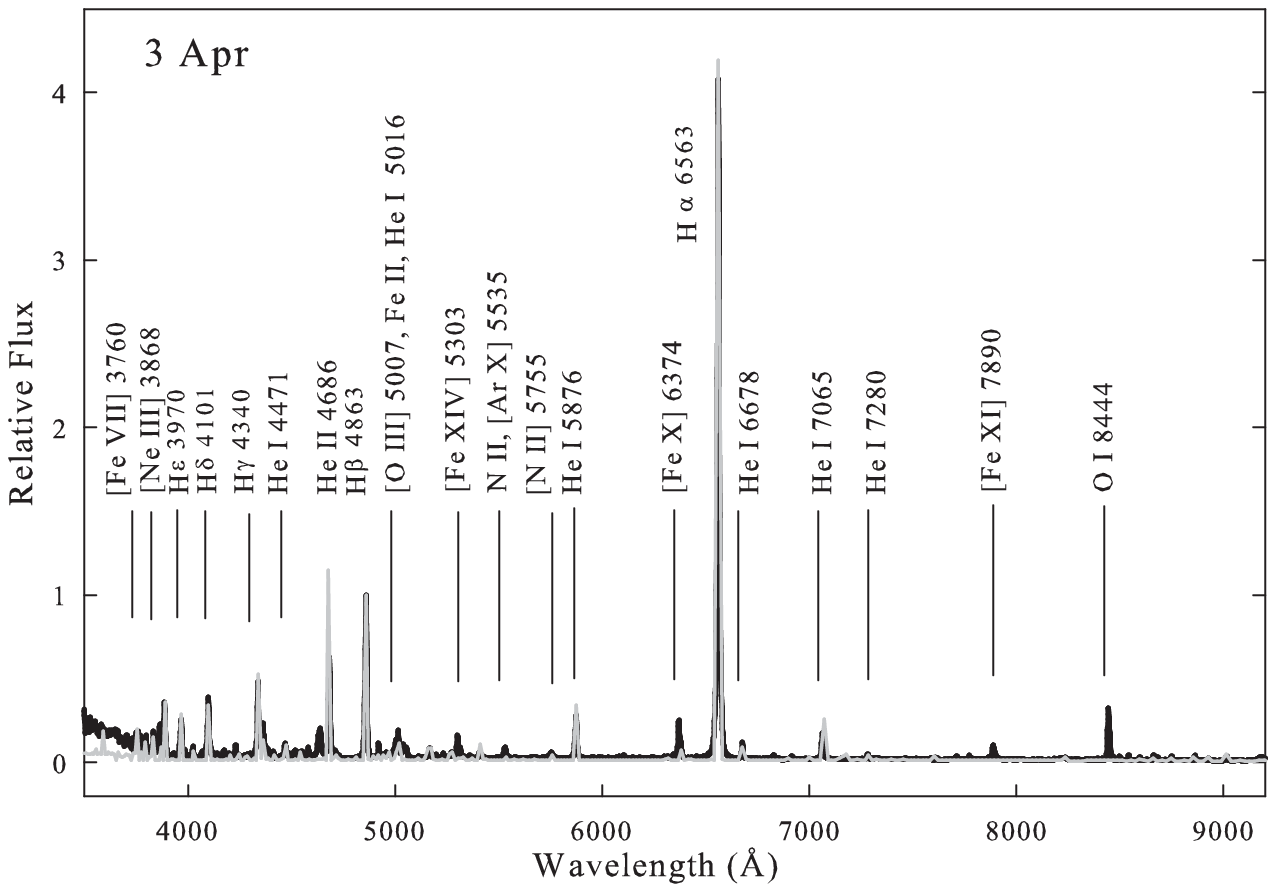}
\caption{Best-fit CLOUDY model spectra (grey line) plotted over the observed spectra (black lines) of RS Oph observed on 2006 April 3.
The spectra were normalized to H$\beta$. Also a few of the strong features have been marked (see text for details).}
\end{figure}

\begin{figure}
\label{cloudy26May}
\centering
\includegraphics[width= 3.3 in, height = 2.65 in, clip]{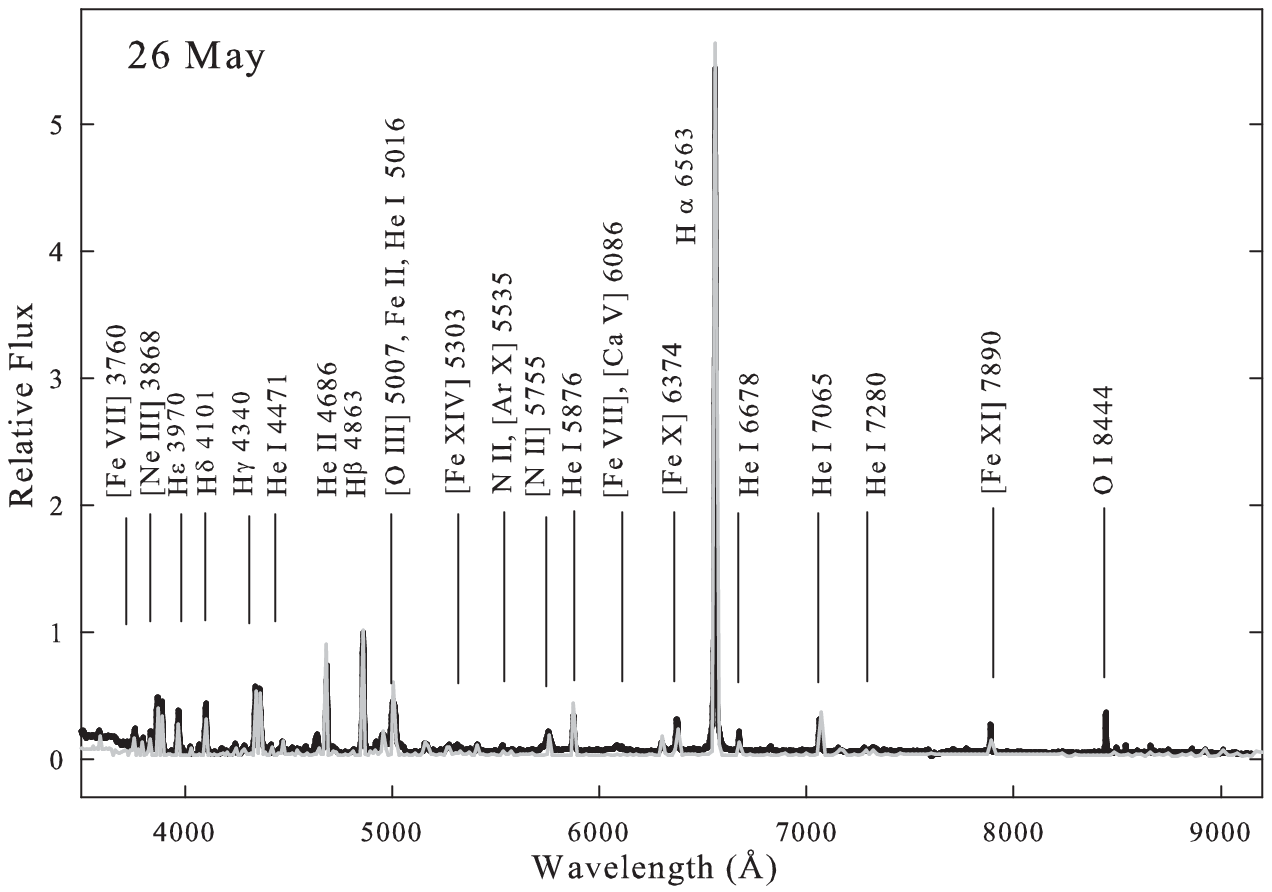}
\caption{Best-fit CLOUDY model spectra (grey line) plotted over the observed spectra (black lines) of RS Oph observed on 2006 May 26.
The spectra were normalized to H$\beta$. Also a few of the strong features have been marked (see text for details).}
\end{figure}

\begin{table*}
\label{cloudyJulOct}
\centering
\caption{Observed and best-fit CLOUDY model line fluxes.}
\smallskip
\centering
\begin{tabular}{c c c c c ccccccccccccc}
\hline
\hline
$\lambda$ &	      &      &  Jul. 19   &            &       &	 Aug. 18  &	 	       &      &  Oct. 17  &   \\
 (\AA)  & Line  & Obs. &  Mod.    & $\chi^{2}$ & Obs.  &  Mod.     & $\chi^{2}$ & Obs.  &  Mod.     & $\chi^{2}$  \\
\hline
3868    & [Ne III] 		 		& 0.80 &	0.65&1.00	& 0.44 &	0.51	&0.22	& 0.30 &0.54&2.56\\
3889    & H$\zeta$, He I  		& 0.38 &	0.29	&0.36	& 0.26 &	0.18	&0.28	& 0.28 &0.20&028\\
3970    & H$\epsilon$, [Ne III] 	& 0.43 &0.31&0.64	& 0.37 &	0.19	&1.44	& 0.30 &0.18&0.64\\
4026    & He I, He II  			& 0.09 &0.04	&0.11	& 0.08 &	0.02	&0.16	& 0.08 &0.02&0.16\\
4101    & H$\delta$   			& 0.38 &0.28	&0.44	& 0.43 &	0.28	&1.0	0	& 0.31 &0.27&0.07\\
4180    & Fe II, [Fe II] 		& 0.06 &	0.02	&0.07	& 0.09 &	0.01	&0.28	&...&...&...\\
4244    & [Fe II] 				& 0.12 &	0.10	&0.02	& 0.16 &	0.02	&0.87	& 0.23 &0.02&1.96\\
4340    & H$\gamma$       		& 0.56 &	0.46	&0.44	& 0.49 &	0.49	&0.00	& 0.34 &0.43&0.36\\
4363    & [O III] 				& 0.72 &	0.54	&1.44	& 0.31 &	0.32	&0.00	& 0.09 &0.05&0.07\\
4415    & [Fe II] 				& 0.09 &	0.08	&0.00	& 0.08 &	0.01	&0.22	& 0.14 &0.02&0.64\\
4471    & He I 	          		& 0.08 &	0.09	&0.00	& 0.11 &	0.08	&0.04	& 0.14 &0.04&0.44\\
4686    & He II 					& 0.65 &	0.66	&0.00	& 0.62 &	0.60&0.02	& 0.29 &0.50&1.96\\
4863		& H$\beta$				& 1.00 &1.00& 0.00  & 1.00& 1.00&0.00	& 1.00& 1.00& 0.00\\
4922    & Fe II, He I			& 0.10 &	0.0	&0.28	& 0.14 &	0.02	&0.64	& 0.13 &0.01&0.64\\
4959    & [O III] 				& 0.94 &	0.89	&3.00	& 0.45 &	0.37	&0.28	& 0.19 &0.22&0.04\\
5007    & [O III], Fe II, He I	& 2.00 &	1.62	&6.42	& 1.18 &	0.82	&5.76	& 0.89 &0.67&2.15\\
5156    & [Fe II], [Fe VII]  	& 0.13 &	0.08	&0.11	& 0.13 &0.02	&0.54	& 0.12 &0.02&0.44\\
5755    & [N II] 				& 0.35 &0.41&0.16   & 0.12 &	0.15	&0.04	& 0.38 &0.40	&0.02\\
5876    & He I        			& 0.28 &	0.30	&0.02	& 0.25 &	0.15	&0.44	& 0.43 &0.23&0.54\\
6086    & [Fe VII], [Ca V] 		& 0.13 &	0.30	&1.28	&...&...&...&...&...&...\\
6300    & [O I], [S III] 		& 0.26 &0.48	&2.15	&...   &... &...    & 0.15 &	0.04	&0.54\\
6374    & [Fe X]			 		& 0.17 &	0.02	&1.0		& 0.21 &	0.04	&1.28	&...&...&...\\
6563    & H$\alpha$   			& 5.12 &5.21	&0.36	& 3.61 &	3.73	&0.64	& 4.03 &4.06&0.04\\
6678    & He I        			& 0.11 &	0.07	&0.07	& 0.13 &	0.03	&0.44	& 0.13 &	0.03	&0.4\\
7065    & He I		  			& 0.16 &	0.18&0.02	& 0.14 &0.10&0.07	&...&...&...\\
7155    & [Fe II] 				& 0.05 &	0.02	&0.02	& 0.03 &	0.02	&0.00	&...&...&...\\
\hline
Total &							&...	   &...	&19.44	& ...  &... &14.69	&...&...	&15.24\\
\hline
\end{tabular}
\end{table*}

\subsection{Nebular Phase: 2006 July -- October}

Prominent appearance of higher ionisation lines like [O III](4363, 4959, 5007 
\AA) and [O I] (6300 \AA) in the day 103 spectrum indicates
that the nova is entering the nebular phase. Due to the expansion of the shell, 
the cloud becomes thinner and consequently higher ionisation lines are seen
in the spectra. Modelling of nebular phase was simpler as only one component 
of relatively lower hydrogen density was sufficient to generate the high as
well as the low ionisation emission lines observed in the spectra.
The line fluxes and $\chi^{2}$ values are presented in Table 3
and the best-fit model parameters are presented in Table 4.
During this phase, the effective temperature and luminosity were increased to 
10$^{5.95}$ K and 10$^{37.5}$ ergs s$^{-1}$, respectively, and the hydrogen 
density was decreased to 10$^{7.18}$ cm$^{-3}$ to match the observed spectra. 
The estimated value of temperature is in agreement with the WD temperature of 
around $8\times 10^{5}$ K derived from X-ray studies (Nelson et al.\ 2008).
During the early phase, we have seen that the abundances of several elements 
were enhanced relative to solar, but as the nebular phase set in, the oxygen, 
neon and iron abundances become sub-solar, whereas, sulphur abundance enhanced 
from a solar abundance seen in the early phase. In the nebular phase also the 
calcium abundance, which was at solar value, was calculated using only the 
blended Ca V (6086 \AA) line. But, the nitrogen abundance was determined 
correctly as the prominent [N II] (5755 \AA) line is highly sensitive to the 
change in abundance parameter in CLOUDY. The best fit modelled spectra (grey 
lines) with  the observed optical spectra (black lines) during this phase are 
shown in Figures 23 - 25.

\begin{table*}
\label{cloudyfit}
\caption{Best-fit CLOUDY model parameters during outburst phase (2006).}
\smallskip
\begin{threeparttable}
\centering
\begin{tabular}{l c c c c c c c ccccc}
\hline
\hline
 Parameters                   &	 Feb.   &	Feb.  	& Mar. 		&  Apr.	 	&   May  	& Jul  		& Aug 		&  Oct. \\
                              &     21   &  24  		& 12   		&   3   		&    26  	&  19  		& 18  		&  17    \\
							 &   (D9)$^{a}$	&	(D12)	&	(D28)	&	(D50)	&	(D103)	&	(D159)	&	(D189)	&(D249)\\
\hline\\	
Log($T_{BB}$) (K)             & 4.45 	& 4.5 		& 5.0 		& 5.5   		&  5.6 		&  5.8  		& 5.9    	&  5.95        \\
Log(luminosity) (erg s$^{-1}$)& 36.9     & 36.8  	& 37 		& 37  		&  37.1 		&  37   		&  37.5   	&   37.3       \\
Log(Hden) (cm$^{-3})$ 		 &	...		&	...		&	...		&	...		&	...		&	7.74		&	7.61	 	&    7.18		\\
Log(Hden)[Clump] (cm$^{-3})$  &   11.0   &  10.8 	& 9.5   		& 8.8   		&  8.5   	&  ...  		&    ...    	&   ...     		\\
Log(Hden)[Diffuse] (cm$^{-3})$&    10.5  &   9.8 	& 8.2   		& 8.0   		&  7.6   	&   ... 		&  ...      	&    ...			 \\
Clump to diffuse covering factor& 70/30  &  70/30	& 80/20 		& 85/15 		&  90/10 	&  	...		&	...	 	&		...	\\
$\alpha^{b}$	          &   -3     &   -3		& -3	    		&  -3		& -3		 	&  -3  		& -3      	&   -3 		 \\
Log(R$_{in}$) ( cm )			  & 	  13.94  & 14.01 	& 14.28 		& 14.45 		& 14.71  	&  15.0		& 15.1  		&  15.28      \\
Log(R$_{out}$) ( cm)          &   14.26  &  14.33 	& 14.63 		& 14.83 		& 15.13  	& 15.3 		& 15.4    	&  15.6       \\
Filling factor	              &	  0.1	 &  0.1  	& 0.1   		& 0.1   		&  0.1   	&  0.1 		&  0.1    	&   0.1    \\
$\beta^{c}$		          &	0.0	     &  0.0  	& 0.0  		& 0.0   		&  0.0   	&  0.0 		&  0.0    	&   0.0    \\
He/He$_\odot^{d}$        &   1.7(9)  &  1.75(10) & 1.8(10)	&  1.9(11) 	&  2.5(11)  	&  2.0(8)   	&  1.6(8)   	&    1.6(6)  \\
N/N$_\odot$                   &   ... 	 &  ...     & 11.0(1)	&  12(3) 	&  10(3)    	&  5.0(1)   &  8.0(1)	&    3.0(1)  \\
O/O$_\odot$                   &   ...	 &  ...		& 1.0(1)		&  1.0(2) 	&  5(3)    	&  0.7(5)   	&  0.3(4)   	&    0.3(4)  \\
Ne/Ne$_\odot$		         &   1.9(1)  &  2.0(1)	& 2.0(1)		&  2.5(2) 	&  5(3)    	&  1.0(2)   	&  0.7(2)   	&    0.5(2)  \\
Ar/Ar$_\odot$                 &   ...    &  4.0(1)	& 4.9(2)		&  5.2(2) 	&  5.5(1)   	&  ...    	&  ...   	&    ...  \\
Fe/Fe$_\odot$				 &   2.7(7)  &  2.8(7)	& 3.0(4)		&  3.5(9) 	&  3.8(8)   	&  0.5(9)   	&  0.5(8)   	&    0.5(6)  \\
Ca/Ca$_\odot$                 &   ...    &  ...		& ...		&  ... 		&  1.0(1)   	&  1.0(1)  	&  ...   	&    ...  \\
S/S$_\odot$                   &   ...    &  ...		& ...		&  ... 		&  1.0(1)   	&  1.5(1)   	&  ...	  	&     1.5(1)  \\
Ni/Ni$_\odot$                 &   1.5(1) &  1.8(1)	& 2.0(1)		&  2.0(1) 	&  2.0(1)   	&  ...   	&  ...   	&   ...  \\
CaseB-$\tau$-Value			 &		4.8	&	4.2	   &		...     &	...	    &		...	&		...	&	...		&	...		\\
Number of observed lines (n)  &    27    &   28     & 26         &  36		&  38	    &   26		&  24		&   21		\\
Number of free parameters (n$_{p}$)&  9   &   10     &  12        & 	12		& 	14      &   11		&   10	    &   10		 \\
Degrees of freedom ($\nu$)		  &  18  &	18     &	  14	        &   24		&   24		&   15	     &   14     &    11		\\
Total $\chi^{2}$ 	              & 21.60&	20.28  &	  17.71     & 30.49      & 32.94		&   19.44    &   14.69  &  15.24 	\\
$\chi^{2}_{red}$                  & 	1.2	 &	 1.13  &  1.26		& 1.27		& 1.37      &   1.29     &   1.05	&   1.39	 \\
\\
\hline
\end{tabular}
\begin{tablenotes}
\item{a} Time elapsed after discovery (in days)
\item{b} Radial dependence of the density $r^{\alpha}$
\item{c} Radial dependence of filling factor $r^{\beta}$
\item{d} Abundances are given in logarithmic scale, relative to hydrogen. All other elements which are not listed
in the table were set to their solar values. The number in the parentheses represents number of lines used in
determining each abundance.
\end{tablenotes}
\end{threeparttable}
\end{table*}

\begin{figure}
\label{cloudy19Jul}
\centering
\includegraphics[width= 3.3 in, height = 2.65 in, clip]{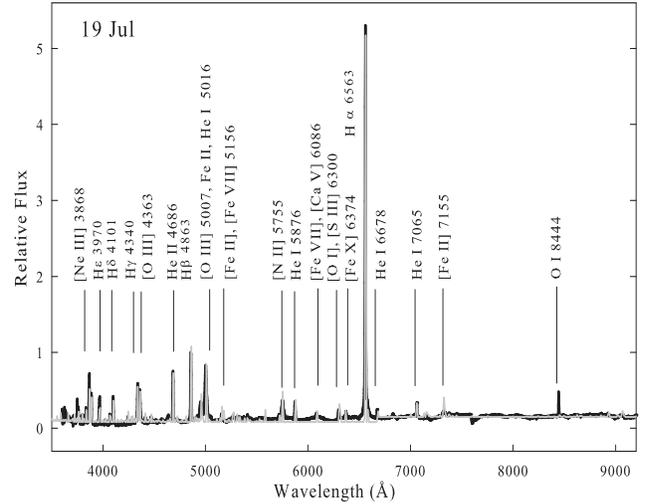}
\caption{Best-fit CLOUDY model spectra (grey line) plotted over the observed spectra (black lines) of RS Oph observed on 2006 July 19.
The spectra were normalized to H$\beta$. Also a few of the strong features have been marked (see text for details).}
\end{figure}

\begin{figure}
\label{cloudy18Aug}
\centering
\includegraphics[width= 3.3 in, height = 2.65 in, clip]{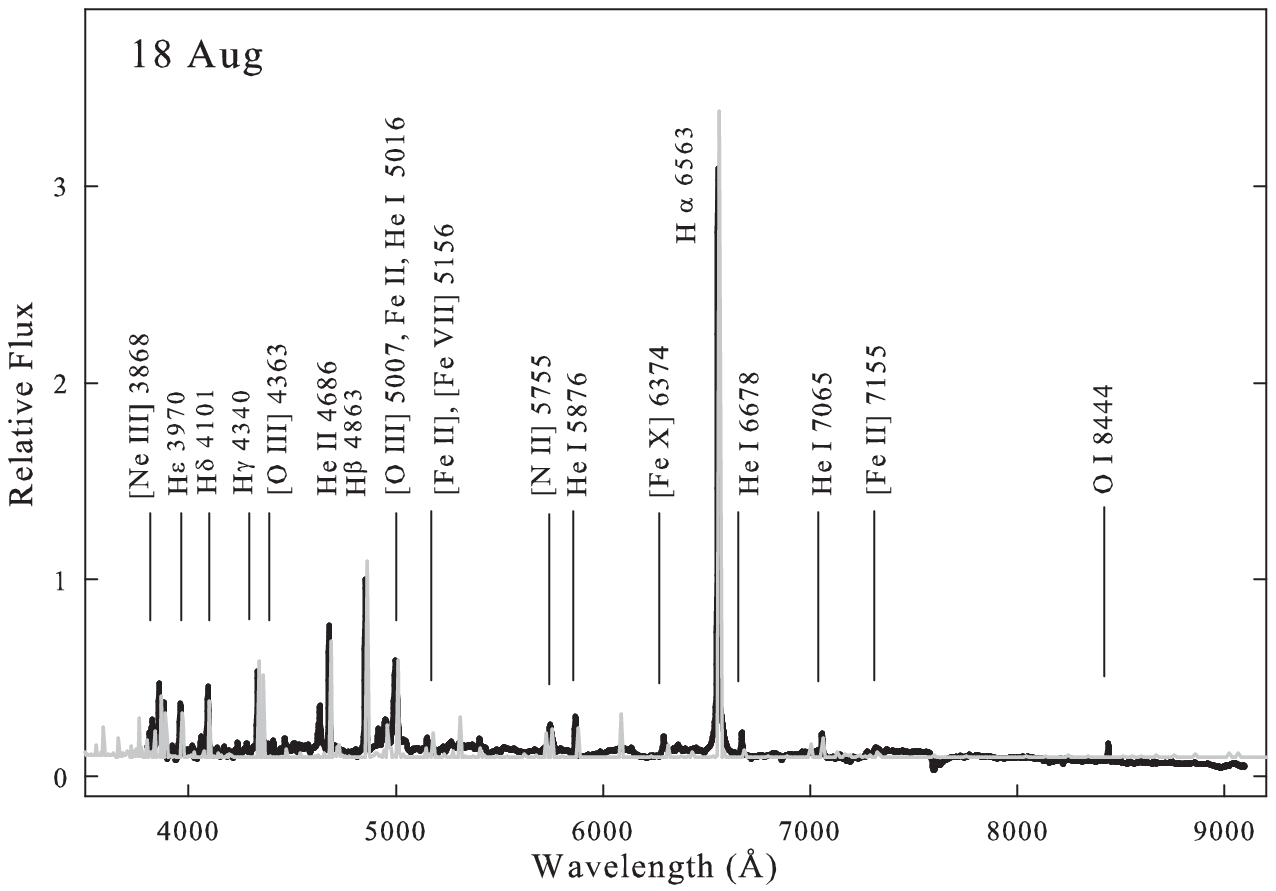}
\caption{Best-fit CLOUDY model spectra (grey line) plotted over the observed spectra (black lines) of RS Oph observed on 2006 August 18.
The spectra were normalized to H$\beta$. Also a few of the strong features have been marked (see text for details).}
\end{figure}

\begin{figure}
\label{cloudy17Oct}
\centering
\includegraphics[width= 3.3 in, height = 2.65 in, clip]{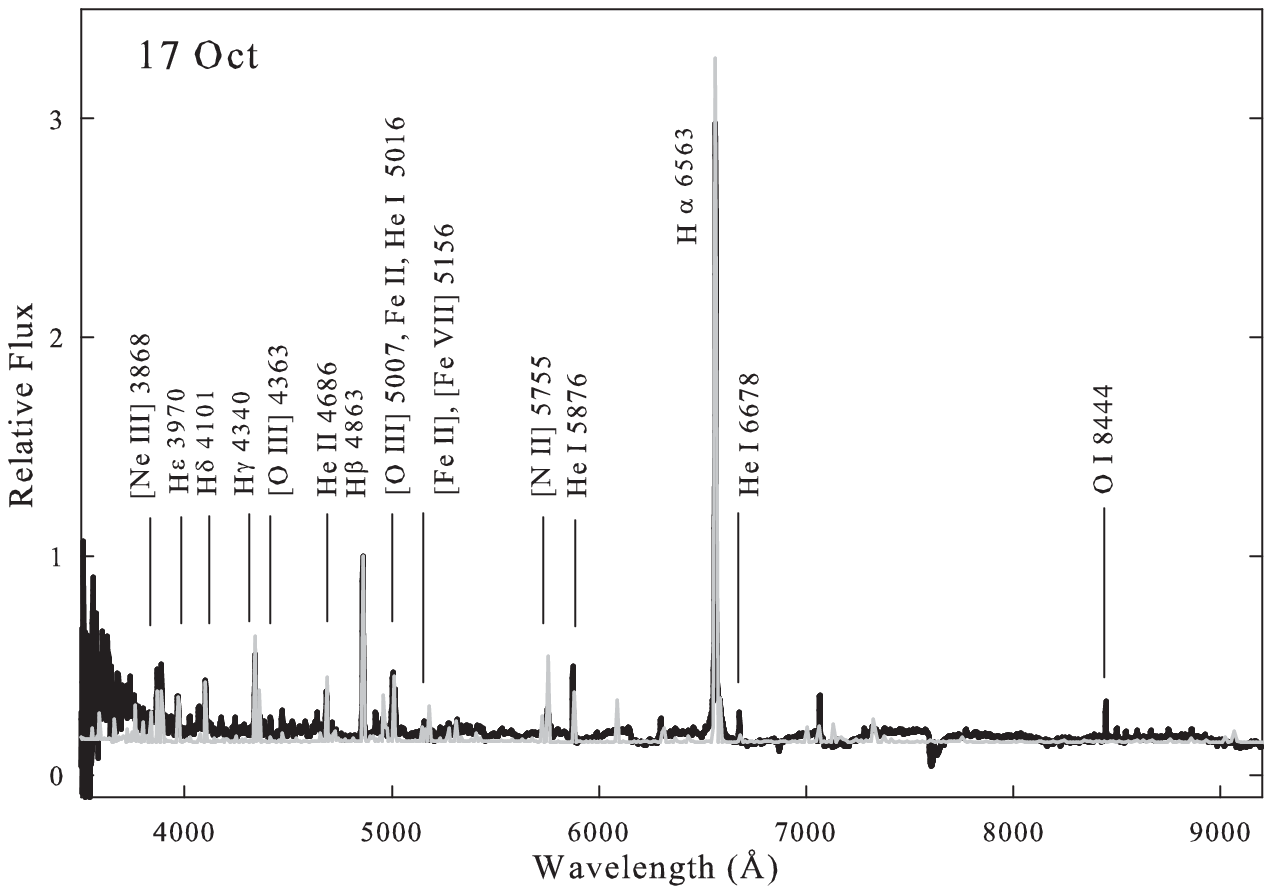}
\caption{Best-fit CLOUDY model spectra (grey line) plotted over the observed spectra (black lines) of RS Oph observed on 2006 October 17.
The spectra were normalized to H$\beta$. Also a few of the strong features have been marked (see text for details).}
\end{figure}

\subsection{Quiescence Phase}

On 17 October, around $\sim 250$ days after the 2006 nova outburst, the spectrum
shows a red continuum with clear TiO absorption bands at 4762, 5448, 5598, 5629 
and 6159 \AA. The absorption band indices indicate the secondary spectral type 
to be M2-M3. By this day the nebular line strengths have decreased. The Fe II 
permitted lines normally seen in quiescence begin to appear. A year later, in 
2007 April, when the nova reached quiescence, the continuum is bluer compared 
to the day $\sim 250$ spectrum, indicating the presence of an accretion disk. 
The quiescence phase spectrum shows prominent emission lines of H and He I, 
together with strong TiO absorption features that indicate the dominance of the 
secondary in the spectrum.

The elemental abundances during quiescence period was estimated by 
photoionisation modelling of the spectrum observed on 2007 April 26.
To generate the model spectrum, three components were considered: a WD 
(T = $10^{4.2}$ K) as an ionising source, a cylindrical accretion
disk (\textit{semi height} = $10^8$ cm) surrounding the source and the secondary
red giant of spectral type M2 III (Worters et al. 2007 and
Anupama \& Mikolajewska 1999). The final spectrum was obtained by adding the 
fluxes generated by these three components. The relative fluxes of the
best-fit model predicted lines, observed lines and corresponding $\chi^{2}$ 
values during quiescence period are presented in Table 5, and
the best fit model parameters are presented in Table 6.
The best fit parameters lead us to an accretion disk of low temperature (10$^{4.2}$ K) 
and low luminosity (10$^{33}$ ergs s$^{-1}$, and a luminosity of 
10$^{35}$ ergs s$^{-1}$ for the red giant). As the dense H-rich material is 
accreted to the disk, the hydrogen density of the disk (10$^{11}$ cm$^{-3}$) is
$\sim 10$ times higher than that in the red giant (10$^{10}$ cm$^{-3}$). The 
$R_{in}$ and $R_{out}$ of the disk are calculated considering a high mass WD 
of 1.35 M$_{\odot}$(e.g.\ Kato, Hachisu \& Luna 2008). During the quiescence 
phase, the observed spectrum shows prominent H and He lines only. The estimated 
abundances of helium is 0.8, which is subsolar. Since there were no prominent 
lines of other elements, the abundances of other elements could not be 
estimated. The best fit modelled spectrum (grey lines), along with the observed 
optical spectrum (black lines) during this phase is shown in Figure 26.

\begin{figure}
\label{cloudyQui}
\centering
\includegraphics[width= 3.3 in, height = 2.65 in, clip]{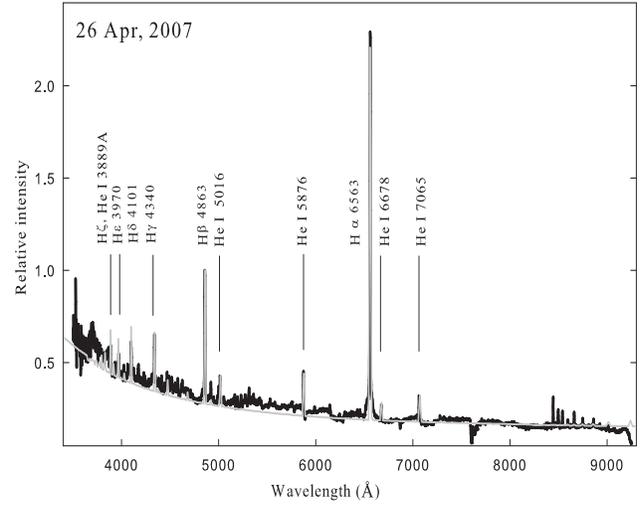}
\caption{Best-fit CLOUDY model spectra (grey line) plotted over the quiescence phase spectra (black line) of RS Oph observed on 2007 April 26.
The spectra were normalized to H$\beta$. Also a few of the strong features have been marked (see text for details).}
\label{quies_cloudy}
\end{figure}

\begin{table}
\label{cloudyApr2007}
\centering
\caption{Observed and best-fit CLOUDY model line fluxes on 2007 Apr 26 (D438).}
\smallskip
\centering
\begin{tabular}{c c c c c }
\hline
\hline
$\lambda$ &	      				&      	&  26 Apr 2007   &\\
 (\AA)  & Line  				& Obs. 	&  Mod.    		& $\chi^{2}$\\
\hline
3889    & H$\zeta$, He I  		& 0.23 	&	0.27		&	0.16\\
3970		& H$\eta$				& 0.20	&   0.22		&   0.04\\
4101    & H$\delta$   			& 0.38 	&	0.35		&	0.11\\
4340    & H$\gamma$       		& 0.34	&	0.38		&	0.13\\
4863		& H$\beta$				& 1.00 	& 	1.00		& 	0.00\\
5016    & He I  					& 0.27 	&	0.25		&	0.01\\
5876		& He I					& 0.45	&	0.39		&	0.36\\
6563    & H$\alpha$   			& 3.60	&	3.47		&	1.72\\
6670    & He I        			& 0.13  &	0.19		&	0.36\\
7065    & He I		  			& 0.18  &	0.20		&	0.01\\
\hline
Total &							&	...	& 	  ...   & 	2.89 \\
\hline
\end{tabular}
\end{table}

\begin{table}
\label{cloudyfitApr2007}
\caption[]{Best-fit CLOUDY model parameters during the quiescence phase}
\smallskip
\begin{threeparttable}
\centering
\begin{tabular}{l c c c c c c c}
\hline\noalign{\smallskip}
 Parameters                   &	 Acc. Disk(35\%)/   \\
                              &  Red Giant (65\%)\\
\hline\noalign{\smallskip}
Log($T_{BB}$) (K)             &   4.2 / 4.2    \\
Log(luminosity) (erg s$^{-1}$)&   33 / 35    \\
Log(Hden) (cm$^{-3})$ 		  &	  11 / 10	\\
$\alpha^{a}$	          &   -3 / -3 \\
Log(R$_{in}$) ( cm )		  &    9 / 12   \\
Log(R$_{out}$) ( cm)          &   11 / 15    \\
Filling factor	              &	 0.1 / 0.1    \\
$\beta^{b}$		      &	 0.0 / 0.0    \\
He/He$_\odot^{c}$        &  0.8(4) / 0.8(4) \\
Number of observed lines (n)  &    10      \\
Number of free parameters (n$_{p}$)&  8   \\
Degrees of freedom ($\nu$)		  &  2   \\
Total $\chi^{2}$ 	              & 2.89  	\\
$\chi^{2}_{red}$                  & 1.44 \\
\hline
\end{tabular}
\begin{tablenotes}
\item{a} Radial dependence of the density $r^{\alpha}$
\item{b} Radial dependence of filling factor $r^{\beta}$
\item{c} Abundances are given in logarithmic scale, relative to hydrogen. All other elements which are not listed
in the table were set to their solar values. The number in the parentheses represents number of lines used in
determining each abundance.
\end{tablenotes}
\end{threeparttable}

\end{table}

\subsection{Comments on the models}

Low values of $\chi^{2} _{red}$ (between 1.05 to 1.44) indicate that the model generated spectra match well with the observed ones. However, inspite of
the low $\chi^{2} _{red}$ values, there are a
few discrepancies. An inspection of Tables 2 and 3
shows that $\chi^{2}$ values of some lines are relatively higher.
For example, H$\alpha$ (6563 \AA) on most of the epochs and H$\gamma$ (4340 \AA) on 2006 February 21 are the highest contributors to
the $\chi^{2}$ values.
Hence, total $\chi^{2}$ value increases when these two strong lines are considered.
Also, while adding the two components, emphasis was given to minimize
the total $\chi^{2}$ value, which leads to difficulty in matching the line flux ratios perfectly to the observed one. This problem disappears during
the nebular phase spectra modelling where only one component of Hydrogen density was enough to fit all the lines. The other lines that contribute to
higher $\chi^{2}$ values are: [Fe X] (6374 \AA) and He I (6678 \AA) on 2006 
April 3. This is not surprising since [Fe X] is a coronal line, most likely 
excited by the shock interaction. The He I line is suppressed due to a high optical depth, which is not well accounted for in the model. 
The He II line at 4686 \AA\ in the modelled spectrum was weak on February 21 and 24. This is possibly because the temperature of the source on those two epochs ($10^{4.45}$ \& $10^{4.5}$\ K respectively) was low.  
Higher temperature $\ga 10^{4.7}$\ K is required to generate the He II line (ionization potential $ \sim 54.4$\ eV) prominently, as seen in the spectra of March 12 onwards, 
when the central source temperature was $\ga 10^5$\ K. Further, the He II 4686 line was very weak in the observed spectra on those two epochs, and it was difficult to measure the flux correctly. Therefore, the He II feature was not included in the calculation of the total $\chi^{2}$ for those epochs.
In spite of the discrepancies, these fits do minimize the total $\chi^{2}$ 
values. During the nebular phases, the highest contributions to the
total $\chi^{2}$ is due to the blending of [O III], Fe II and He I lines at 
5007 \AA. Also, the abundances for Fe, He and O were set after matching the 
other Fe, He and O lines, which leads to a high (about 30 - 40$\%$) $\chi^{2}$ 
contribution. Additionally, the model could not generate a few of the observed 
lines, e.g., N III (4640 \AA), Si II (5041, 5056 \AA), Ca II (8668, 8548 \AA), 
O I (7774, 8446 \AA) and O VI (6106 \AA). Some of these issues have been 
reported in our previous CLOUDY analysis of RS Oph spectra (Das \& Mondal, 
2015). So, these lines have been excluded while calculating the values of 
$\chi^{2}$. This is possibly due to the reason that even a two-component model 
can not properly represent the complex density structure of the nova ejecta.

\section{Summary}
\begin{enumerate}
\item{}
This is the first outburst of RS Oph which has revealed the complexity of emission mechanisms. This has been possible only due to the wide spectral and temporal coverage of the observations.
\item{}
The evolution of the optical spectrum is very similar to previous outbursts.
\item{}
The emission line velocity evolution indicates that the remnant was in the adiabatic phase for a very brief period and quickly moved to the radiative
cooling phase by $t\sim 5$ days. The nova remnant probably reached the edge of the red giant wind by $\sim 80$ days, as indicated by the marginal increase
in the emission line velocity. These findings are consistent with those 
based on X-ray observations (Bode et al. 2006).
\item{}
The nebular lines originate in a region closer to the WD while the recombination lines originate in the decelerating material.
\item{}
The observed spectra are modelled using the photoionisation code CLOUDY for 9 epochs. Based on the modelling, the elemental abundances and other parameters such as the temperature and density of the ejecta, and the luminosity of the WD at the various epochs are estimated.
\end{enumerate}

\section{acknowledgements}
We thank the HCT Time Allocation Committee for a generous allocation of ToO
time. We also thank all the observers of both the HCT and the VBT, who spared
part of their observing time for the nova observations, without which the dense
monitoring during the early phases of the outburst would not have been possible.


\begin{thebibliography}{99}

\bibitem[Adams $\&$ Joy(1933)]{adams}Adams W. S. $\&$ Joy A. H., 1933, PASP, 45, 301
\bibitem[Anupama $\&$ Mikolajewska(1999)]{anupama}Anupama G. C. $\&$ Mikolajewska J., 1999, A$\&$A, 344, 177
\bibitem[Anupama $\&$ Prabhu(1989)]{anupama1} Anupama G. C. $\&$ Prabhu T. P., 1989, JAp$\&$A, 10, 237
\bibitem[Banerjee et al. (2009)]{BDA} Banerjee D.P. K., Das R. K. $\&$ Ashok N. M ., 2009, MNRAS, 399, 357
\bibitem[Bode $\&$ Kahn(1985)]{bode}Bode M. F. $\&$ Kahn F. D., 1985, MNRAS, 217, 205-215
\bibitem[Bode(1987)]{bode1} Bode M. F.,  1987, Proceedings of the meeting on RS Ophiuchi (1985) and the Recurrent Nova Phenomenon, Utrecht: VNU Science Press, (Ed: Bode, M.F)
\bibitem[Bode et al. (2006)]{bode2} Bode M. F., O'Brien T. J., Osborne J. P. et al. 2006, ApJ, 652, 629
\bibitem[Buil(2006)]{buil} Buil C., 2006, Cent. Bur. Electron. Tel., 403, 1
\bibitem[Das et al. (2006)]{das} Das R. K., Banerjee D. P. K. $\&$ Ashok N. M., 2006, ApJ, 653, L141
\bibitem[Das $\&$ Mondal(2015)]{das1} Das R. K. $\&$ Mondal A., 2015, New Ast., 39, 19
\bibitem[Evans et al.(2007)]{evans} Evans A., Kerr T., Bin Yang et al,. 2007, MNRAS, 374, L1
\bibitem[Eyres et al. (2009)]{Eyres} Eyres S. P. S., O'Brien T. J., Beswick R. et al., 2009, MNRAS, 395, 1533
\bibitem[Ferland et al.(2013)]{ferland} Ferland G. J., Porter R. L., van Hoof P. A. M. et al., 2013, RMxAA, 49, 137
\bibitem[Fujii(2006)]{fujii}Fujii M., 2006, VSNET Alert 8869
\bibitem[Gorbatskii(1972)]{Gorbatskii}Gorbatskii V. G., 1972, Soviet Astronomy, 16, 32
\bibitem[Gorbatskii(1973)]{Gorbatskii}Gorbatskii V. G., 1973, Soviet Astronomy, 17, 11
\bibitem[Helton et al.(2010)]{helton} Helton L. A., Woodward C. E., Walter F. M. et al., 2010, ApJ, 140, 1347
\bibitem[Iijima(2006)]{iijima}Iijima T., 2006, IAU Circ., 8675, 1
\bibitem[Joy $\&$ Swings(1945)]{joy}Joy A. H. $\&$ Swings P., 1945, ApJ, 102, 353J
\bibitem[Josafatson $\&$ Snow(1987)]{josafatson}Josafatson K. $\&$ Snow T. P., 1987, ApJ, 319, 436
\bibitem[Kantharia et al.(2007)]{kantharia} Kantharia N. G., Anupama G. C., Prabhu T. P., Ramya S., Bode M. F., Eyres S. P. S., O'Brien T. J., 2007, ApJ, 667, L171
\bibitem[Kato et al.(2008)]{kato} Kato M., Hachisu I. $\&$ Luna G. J. M., 2008, in RS Ophiuchi (2006) and the
Recurrent Nova Phenomenon, Astronomical Society of the Pacific Conference Series, eds. Evans A., Bode M. F., O$'$Brien T. J. $\&$ Darnley M. J., vol. 401, 308
\bibitem[Monnier et al.(2006)]{monnier} Monnier J. D., Barry R. K., Traubet W. A. et al., 2006, ApJ, 647, L127
\bibitem[Narumi et al.(2006)]{narumi} Narumi H., Hirosawa K., Kanai K., Renz W., Pereira A., Nakano S., Nakamura Y., Pojmanski G., 2006,  IAUC 8671
\bibitem[Nelson et al.(2008)]{nelson} Nelson T., Orio M., Cassinelli J. P. et al., 2008, ApJ, 673, 1079
\bibitem[O$'$Brien et al.(1987)]{brien} O$'$Brien T. J. $\&$ Kahn F. D., 1987, MNRAS, 228, 277
\bibitem[O$'$Brien et al.(2006)]{brien1} O$'$Brien T. J., Bode M. F. $\&$ Porcas R. W. et al., 2006, Nature, 442, 279
\bibitem[Oppenheimer $\&$ Mattei(1993)]{oppenheimer} Oppenheimer B. D. $\&$ Mattei J. A., 1993, JAAVSO, 22, 105
\bibitem[Osborne et al.(2011)]{osborne} Osborne J. P., Page K. L., Beardmore A. P. et al., 2011, ApJ, 727, 124
\bibitem[Rosino(1987)]{rosino} Rosino L., 1987, in  Proceedings of the meeting on RS Ophiuchi (1985) and the Recurrent Nova Phenomenon, Ed: M.F. Bode. Utrecht: VNU Science Press,  p1
\bibitem[Rosino $\&$ Iijima(1987)]{rosino1} Rosino L. $\&$ Iijima T., 1987, in Proceedings of the meeting on RS Ophiuchi (1985) and the Recurrent Nova Phenomenon, Ed: M.F. Bode. Utrecht: VNU Science Press,  p27
\bibitem[Rupen et al.(2008)]{rupen}Rupen M. P., Mioduszewski A. J. $\&$ Sokolowski J. L., 2008, ApJ, 688, 559
\bibitem[Schaeffer(2004)]{}Schaeffer B. E., 2004, IAUC, 8396
\bibitem[Schmid(1989)]{schmid}Schmid H. M., 1989, A\& A, 211, L31
\bibitem[Schwarz(2002)]{schwarz1}Schwarz G. J., 2002, ApJ, 577, 940
\bibitem[Schwarz et al.(2007)]{schwarz2} Schwarz G. J., Shore S. N., Starrfield S. et al., 2007, ApJ, 657, 453
\bibitem[Shore et al.(1996)]{shore}Shore S. N., Kenyon S. J., Starrfield S. $\&$ Sonneborn G., 1996, ApJ, 456, 717
\bibitem[Skopal et al.(2008)]{skopal}Skopal A., Pribulla T., Buil C. et al., 2008, in RS Ophiuchi (2006) and the Recurrent Nova Phenomenon, Astronomical Society of the Pacific Conference Series, eds. Evans A., Bode M. F., O$'$Brien T. J. $\&$ Darnley M. J., vol. 401, 227
\bibitem[Snijders(1987)]{snijders}Snijders M.A.J., 1987. AP$\&$SS, 130, 243.
\bibitem[Sokoloski et al.(2006)]{sokoloski}Sokoloski J. L., Luna G. J. M., Mukai K., $\&$ Kenyon S. J., 2006,  Nature, 442, 276
\bibitem[Starrfield et al.(1985)]{Starrfield}Starrfield S., Sparks W. M., Truran J. W., 1985,  ApJ, 291, 136
\bibitem[Yaron et al.(2005)]{Yaron}Yaron, O.; Prialnik, D.; Shara, M. M.; Kovetz, A,  2005, ApJ, 623, 398
\bibitem[Vanlandingham et al.(2005)]{vanlandingham}Vanlandingham K. M., Schwarz G. J., Shore S. N. et al., 2005, ApJ, 624, 914
\bibitem[Worters et al.(2007)]{worters}Worters H. L., Eyres S. P. S., Bromage G. E. et al., 2007, MNRAS, 379, 1557
\bibitem{}
\bibitem{}
\end{thebibliography}
\end{document}